\documentclass[singlecolumn, trackchanges]{aastex631}
\hypersetup{linkcolor=red, citecolor=blue, filecolor=magenta, urlcolor=cyan}

\newcommand{\blue}[1]{\textcolor{blue}{#1}}

\usepackage{soul}

\renewcommand{\figurename}{Figure\,}
\renewcommand{\tablename}{Table\,}
\newcommand{\sectionname}{Section\,}
\renewcommand{\appendixname}{Appendix\,}
\def\figref#1{\figurename\,\ref{#1}}
\def\tabref#1{\tablename\,\ref{#1}}
\def\secref#1{\sectionname\,\ref{#1}}
\def\appref#1{\appendixname\,\ref{#1}}
\def\eqref#1{(\ref{#1})}

\usepackage{CJK}

\shorttitle{\textit{Gaia} CMD Massive Star Selection in SMC}
\shortauthors{Nakano et al.}
\graphicspath{{./}{figures/}}

\begin{document}

\title{Evidence of Galactic Interaction in the Small Magellanic Cloud Probed by \textit{Gaia} Selected Massive Star Candidates}

 \correspondingauthor{Satoya Nakano}
 \email{nasatoya@a.phys.nagoya-u.ac.jp}

\author[0009-0001-8036-7296]{Satoya Nakano}
\affiliation{Department of Physics, Graduate School of Science, Nagoya University, Furo-cho, Chikusa-ku, Nagoya 464-8602, Japan}

\author[0000-0002-1411-5410]{Kengo Tachihara}
\affiliation{Department of Physics, Graduate School of Science, Nagoya University, Furo-cho, Chikusa-ku, Nagoya 464-8602, Japan}

\author{Mao Tamashiro}
\affiliation{Department of Physics, Graduate School of Science, Nagoya University, Furo-cho, Chikusa-ku, Nagoya 464-8602, Japan}




\begin{abstract}

We present identifications and kinematic analysis of 7,426 massive ($\mathrm{\geq}8M_{\odot}$) stars in the Small Magellanic Cloud (SMC), using \textit{Gaia} DR3 data. We used \textit{Gaia} ($G_\mathrm{BP}-G_\mathrm{RP}$, $G$) color-magnitude diagram to select the population of massive stars, and parallax to omit foreground objects.
The spatial distribution of the 7,426 massive star candidates is generally consistent with the spatial distribution of the interstellar medium, such as H$\alpha$ and H\,\textsc{i} emission. The identified massive stars show inhomogeneous distributions over the galaxy, showing several superstructures formed by massive stars with several hundred parsecs scale.
The stellar superstructures defined by the surface density have opposite mean proper motions in the east and west, moving away from each other. Similarly, the mean line-of-sight velocities of the superstructures are larger to the southeast and smaller to the northwest. The different east-west properties of the superstructures' proper motion, line-of-sight velocity indicate that the SMC is being stretched by tidal forces and/or ram pressure from the Large Magellanic Cloud to the southeast, thereby rejecting the presence of galaxy rotation in the SMC.

\end{abstract}




\section{Introduction}
\label{sec:1}
The Small Magellanic Cloud (SMC) at $\sim$60~kpc (e.g., \citealp{2015AJ....149..179D}) and the Large Magellanic Cloud (LMC) at $\sim$50~kpc (e.g., \citealp{2014AJ....147..122D}) are the nearest interacting dwarf galaxies to the Milky Way. They are suitable for investigating the effects of low metallicity, weak gravitational potential, and galaxy interactions on star formation and galaxy evolution, and are also useful as an analogue of the hierarchical structure formation of the early universe.
They are valuable for observational studies not only because their close distances allow for the individual stars and molecular clouds to be resolved, but also because the center of the Milky Way does not overlap in the line-of-sight and there is minimal contamination from the Galactic disk, enabling comprehensive observations of the entire galaxy.

In particular, the effects of the interaction are stronger for SMC, which is about an order of magnitude lighter than the LMC \citep{1998ApJ...503..674K, 2004ApJ...604..176S}. In the SMC, the interstellar medium is disturbed by tidal forces and/or ram pressure. The ISM shows a bar structure extending from north to southwest \citep{1972VA.....14..163D} and a wing structure extending east \citep{1940BHarO.914....8S}. The wing is connected to the Magellanic Bridge, which forms the flow of stars and gas into the LMC \citep{1963AuJPh..16..570H, 1985Natur.318..160I}.
There is also the Magellanic Stream \citep{1974ApJ...190..291M}, a flow of neutral hydrogen gas toward the center of the Milky Way Galaxy, which may have been formed as a result of the interaction, mainly by the stripping of large amounts of gas from the SMC \citep{2012ApJ...750...36D}.
The SMC shows a different spatial distribution for each stellar population.
Stars older than $\sim$2~Gyr show a nearly spherically symmetric spatial distribution, while stars of $\sim$100~Myr show the bar structure. Furthermore, younger stars, $\sim$10~Myr old, show overdensities in the northern part of the bar and in the wing structure (e.g., \citealp{2018MNRAS.478.5017R, 2019MNRAS.490.1076E}).
Since the distributions of young stars of $\sim$10~Myr old coincide with the spatial distribution of the ISM, the observed difference in spatial distribution among the stellar populations are interpreted as the result of the last close encounter between the SMC and the LMC $\sim$200~Myr ago \citep{2012ApJ...750...36D}. 
In addition, red clump stars show bimodality in distance, with the distances being $\sim$60~kpc and $\sim$50~kpc in the eastern SMC (e.g., \citealp{2017MNRAS.467.2980S, 2021MNRAS.504.2983T}). For Cepheids and old RR Lyrae stars, the line-of-sight depth in the SMC extends over 20~kpc (e.g.,\,\,\citealp{2017AcA....67....1J, 2017MNRAS.472..808R}), suggesting an elongation along the Milky Way.

In recent years, as instrument performance has improved, the proper motion (PM) of stars in the Magellanic galaxies has been successfully tracked; recent studies in the SMC include \citet{2018ApJ...867L...8O}, which showed that the wing is moving toward the LMC, \citet{2018ApJ...864...55Z}, which showed that the SMC outer regions has PMs radially away, and \citet{2021MNRAS.502.2859N}, which showed that both young and old stars in the bar are moving westward. 
Stellar PM provides information perpendicular to the line-of-sight velocity derived from spectroscopic observations, allowing us to understand three-dimensional kinematics.
In particular, the PMs of stars provide information on the interstellar medium dynamics perpendicular to the line-of-sight \citep{2019ApJ...887..267M}, since young stars, including massive stars, should trace the motions of their parent gas.

In this study, we select non-embedded massive stars ($\mathrm{\geq}8M_{\odot}$) throughout the SMC using \textit{Gaia} Data Release 3 (DR3; \citealp{2016A&A...595A...1G, 2023A&A...674A...1G}) and explore their kinematics based on the \textit{Gaia} PMs.
This study's new insight is to focus on stellar mass rather than stellar age, which has been the focus of most previous studies (e.g., \citealp{2019MNRAS.490.1076E,2021A&A...649A...7G}).
By uniformly selecting massive star candidates across the entire galaxy, we characterize their spatial distribution and kinematics, and identify evidence of galaxy interactions that may have induced the formation of massive stars. The massive star candidates we selected are catalogued and made publicly available to anyone using \textit{Gaia} DR3.

\secref{sec:data} describes the \textit{Gaia} DR3 data used in this study.
In \secref{sec:selection} we describe the selection of massive star candidates using the color-magnitude diagram (CMD) and discuss the validity of the selection.
In \secref{sec:PM} we calculate the internal PMs to show motions of the massive star candidates in the SMC, which would trace motions of the ISM.
We also discuss the properties of the radial velocities (RVs) available for some of the massive star candidates.
In \secref{sec:dbscan} we use clustering analysis to show the effects of galaxy interactions on the internal young stars.
\secref{sec:conclusion} describes conclusions.

In this study, we assume a distance modulus to the SMC of 19.06~mag, corresponding to a distance of 65~kpc.

\begin{figure*}
 \centering
 \includegraphics[width=18cm,clip]{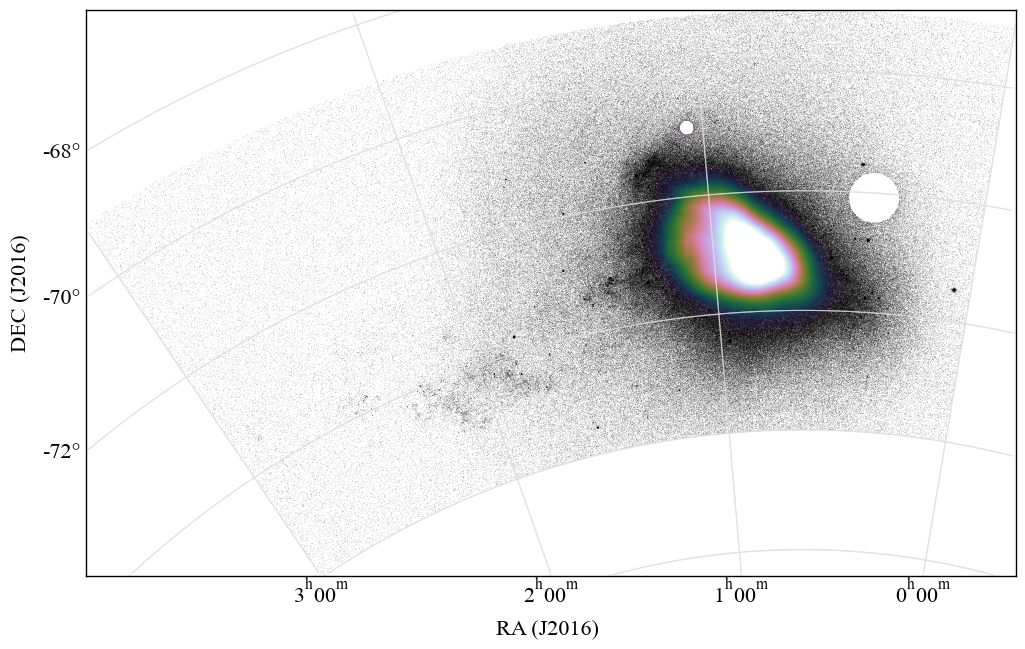}
 \caption{
 Stellar density distribution around SMC obtained from \textit{Gaia} data. Foreground objects are excluded by the limitation of the annual parallax $\varpi >0.1~\mathrm{mas}$ and cuts around two globular clusters NGC~362 and NGC~104.
}
\label{SMC_Gaia}
\end{figure*}

\section{The \textit{Gaia} DR3 Data}
\label{sec:data}

\subsection{SMC Data Selection}
\label{sec:data1}
As SMC region, we obtained \textit{Gaia} DR3 data using a selection with $\mathrm{RA}=00\mathrm{h}\textsc{--}03\mathrm{h}$ and $\mathrm{DEC}=-76^{\circ}$~to~$-69^{\circ}$.
In the obtained data, the brighter stars have more accurate observational quantities; for stars with $G\sim14~\mathrm{mag}$ corresponding to bright massive stars in the SMC, the error in the $G$ magnitude is $\sigma_{G}\approx0.0031~\mathrm{mag}$, the error in the color index $G_\mathrm{BP}-G_\mathrm{RP}$ is $\sigma_{G_\mathrm{BP}-G_\mathrm{RP}}\approx0.0078~\mathrm{mag}$, and the error in the parallax $\varpi$ is $\sigma_{\varpi}\approx0.025~\mathrm{mas}$, however for stars with $G\sim17~\mathrm{mag}$ corresponding to relatively faint massive stars in the SMC, the errors are $\sigma_{G}\approx0.0035~\mathrm{mag}$, $\sigma_{G_\mathrm{BP}-G_\mathrm{RP}}\approx0.0155~\mathrm{mag}$, and $\sigma_{\varpi}\approx0.068~\mathrm{mas}$.
This region includes the main body of the SMC and its surroundings, as well as a part of the Magellanic Bridge.
This region also includes foreground objects that should be removed: stars in the Milky Way and two globular clusters (NGC~362 in the northern part of SMC and NGC~104 in the western part of SMC).
At first, NGC~362 was removed by a cut of radius $7'$ centered on $\mathrm{RA}=01\mathrm{h}\,\,03\mathrm{m}\,\,14.26\mathrm{s}$, $\mathrm{DEC}=-70^{\circ}\,\,50'\,\,55.6''$, and NGC~104 was removed by a cut of radius $25'$ centered on $\mathrm{RA}=00\mathrm{h}\,\,24\mathrm{m}\,\,05.67\mathrm{s}$, $\mathrm{DEC}=-72^{\circ}\,\,04'\,\,52.6''$.

Furthermore, to remove foreground objects in the Milky Way, we excluded objects at $\varpi >0.1~\mathrm{mas}$ (\text{distance $< \mathrm{10~kpc}$}) from the data, without considering errors in $\varpi$.
Because bright red foreground objects that have similar $G$ magnitudes to the SMC massive stars ($G\lesssim14~\mathrm{mag}$) have relatively small errors in the $\varpi$, this limitation on the $\varpi$ allows us to limit the contamination of foreground objects into the SMC massive star samples to about 0.6\% (see \appref{sec:foreground} for details).
Another unique point of our study is that we do not miss high-velocity runaway stars ejected from star clusters by not imposing the PMs constrains on the samples.
However, some stars in the SMC region exhibit truly negative parallaxes ($\varpi <0~\mathrm{mas}$), even after accounting for measurement errors.
Negative parallaxes arise when fitting annual motion from data that include measurement noise, and removing them introduces a significant bias toward nearby sources, which is particularly unsuitable in the analysis of the SMC \citep{2018A&A...616A...9L}.
Therefore, in this study, parallax is used solely as a criterion to determine whether the true value falls below 0.1~mas.
Negative parallaxes can also be attributed to factors such as zero-point offsets \citep{2021A&A...649A...4L}.
These imply that $\varpi$ can also be overestimated, potentially leading to the erroneous exclusion of SMC stars due to the limitation of $\varpi$.
We will discuss that the incompleteness of the massive star sample caused by the limitation of $\varpi$ is sufficiently small in \secref{sec:RecovRate}.
In the context of the false detection rate, our sample of massive star is largely unaffected even without considering the zero-point offsets (see \appref{sec:foreground}).

\figref{SMC_Gaia} shows the final stellar density distribution of the SMC region, excluding foreground sources; the total number of stars in the SMC region is 2,043,092.
Among these, 1,832,516 stars have both $G_\mathrm{BP}$ and $G_\mathrm{RP}$ available.
In our massive star selection method using the CMD, which is described in detail in \secref{sec:CMD}, 
it is a necessary condition that all of $G$, $G_\mathrm{BP}$, and $G_\mathrm{RP}$ are available for inclusion in the selection.

\subsection{Evaluation of Crowding in the SMC}
\label{sec:data2}
In particularly crowded regions, such as young massive clusters, crowding may lead to a degradation in photometric quality and a loss of completeness. In this section, we show that due to the resolution of \textit{Gaia}, the effects of crowding are negligible for the $G$ measurements, while blending caused by crowding affects the $G_\mathrm{BP}$ and $G_\mathrm{RP}$ measurements, potentially leading to decrease in completeness in our catalog.

The spatial resolution of \textit{Gaia} DR3 is typically approximately~$0.7''$ (which corresponds to 0.22~pc at a distance of 65~kpc; stars closer than this distance have a significantly reduced detection rate; \citealp{2021A&A...649A...2L,2021A&A...649A...5F}).
To evaluate the impact of \textit{Gaia} DR3 resolution on crowding of massive stars in the SMC, we use the number of upper main sequence stars in NGC~346.
NGC~346 is a young cluster with several dozen O-type stars (e.g., \citealp{1989AJ.....98.1305M, 2022A&A...666A.189R}), and it is one of the regions with the highest density of massive stars in the SMC; the effect of crowding on massive star selection is the most significant in NGC~346 within the SMC.
\citet{2024ApJ...971..108H} have observed approximately 108,000~$\mathrm{arcsec}^{2}$ of the most active star formation region in NGC~346 using the Near Infrared Camera of the \textit{James Webb Space Telescope}, identifying 3,166 upper main sequence stars.
When 3,166 points are randomly distributed within 108,000~$\mathrm{arcsec}^2$, we obtain an average of 71.1 overlaps (2.24\% of the total) within $0.7''$ after 1,000 iterations.
This rough estimate suggests that, even in NGC~346, only about 2\% of the upper main sequence stars overlap below the resolution limit of \textit{Gaia}.
It is noted that since most of the upper main sequence stars are intermediate-mass stars with masses ranging from $\sim$2--8$M_\mathrm{\odot}$, massive stars are limited in number among them. Even if faint stars overlap with massive stars, the large luminosity difference means their impact on the photometry of massive stars is negligible.
Similarly, \citet{2021A&A...646A.106S} concluded that, based on the well-determined spectral types of the ten brightest objects in the SMC from \textit{Gaia} DR2, these objects are not part of an unresolved cluster containing a large number of bright stars.

Furthermore, if multiple objects are detected within $0.18''$ by \textit{Gaia}, one of them is discarded \citep{2021A&A...649A...4L}, making it impossible to detect (multiple) binary systems in the SMC.
Accurate estimates of the binary fraction in low-metallicity environments such as the SMC are lacking, and the impact of unresolved close binaries on the completeness of the data is unknown. As an early estimate of the binary fraction in massive stars in the SMC, \citet{2024ApJ...972....3K} report an estimate of the wide binary fraction of less than 5\%, suggesting that the binary fraction in the SMC may be significantly smaller than that in the Milky Way.

The window size used for the spectrophotometric prisms using $G_\mathrm{BP}$ and $G_\mathrm{RP}$ measurements is relatively large, measuring $3.5'' \times 2.1''$, which can lead to blending of multiple objects in crowded regions \citep{2021A&A...649A...3R,2021A&A...652A..86C}.
Due to the slight overlap between the $G_\mathrm{BP}$ and $G_\mathrm{RP}$ bands, this blending is evaluated using the photometric excess factor $C$, defined as the sum of the fluxes in $G_\mathrm{BP}$ and $G_\mathrm{RP}$ divided by the flux in $G$ \citep{2018A&A...616A...4E}.
For instance, \citet{2021A&A...649A...7G} utilized the fact that the $C$ is typically below 1.4 for most isolated objects, and demonstrated that in the bar of the SMC, objects with a $C$ exceeding 1.4 are more common compare to the outer regions.
Furthermore, \citet{2021A&A...649A...3R} fitted $C$ as a function of $G_\mathrm{BP}-G_\mathrm{RP}$ for isolated stars with high-quality observations.
They then calculated the $1\sigma$ range of $C^{*}$, the residuals from the fitting result, and proposed sigma-clipping of $C^{*}$ as a method for excluding stars with inconsistent photometry.
Applying the 1$\sigma$ clipping to $C^{*}$ to the 1,832,516 stars in the SMC with available $G_\mathrm{BP}-G_\mathrm{RP}$ results excludes 1,193,759 stars (corresponding to approximately 65\%). 
Thus, applying the $C^{*}$ constraint to the SMC stars is not suitable for maintaining completeness.
This highlights the limitations of the $G_\mathrm{BP}$ and $G_\mathrm{RP}$ measurements in \textit{Gaia} DR3 for the SMC, with improvements expected in \textit{Gaia} DR4.

In contrast, stars satisfying $C > 1.4$, as used by \citet{2021A&A...649A...7G} for crowding assessment, account for about 17\% of the 1,832,516 stars, or 307,084 stars.
Among the stars with $C > 1.4$, only 3,710 stars have $G < 17~\mathrm{mag}$, and only 39 stars have $G < 14~\mathrm{mag}$. The excess flux due to blending is less likely to occur for brighter stars, suggesting that the impact on massive star selection is relatively small.
$C = 1.4$ corresponds to an excess of approximately 20\% in flux for massive stars (see \secref{sec:3.2.2}).
Since the number of adjacent bright blue stars is limited, our sample contains very few intermediate- to low-mass stars that are mistakenly classified as blue massive star candidates due to blending with neighboring blue massive stars, as shown in \secref{sec:3.2.2}.
However, there remains the possibility that true blue massive stars could be excluded from our massive star selection due to blending with adjacent red giants, making their detection challenging.
The number of stars satisfying $C > 1.4$ suggests that the number of stars missed from our catalog due to such blending is likely less than 3,710 stars, even when estimated conservatively. 
The fact that the number of stars in the red giant branch is comparable to that on the upper main sequence in NGC~346 \citep{2024ApJ...971..108H} suggests that the probability of blending between bright red giants and massive stars is even smaller than 2\%.

In the worst case, crowding may lead to a decrease in the completeness of the $G_\mathrm{BP}$ and $G_\mathrm{RP}$ measurements in crowded regions, as the measurements of adjacent bright stars are prioritized \citep{2021A&A...649A...3R}.
To estimate the number of stars for which $G_\mathrm{BP}$ and $G_\mathrm{RP}$ photometry is unavailable due to crowding, we use the 21 O-type stars listed in \citet{2022A&A...666A.189R} from the core of NGC~346.
\citet{2022A&A...666A.189R} provided a nearly complete list of O-type stars in the central $1'$ of NGC~346 by spectroscopically observing this area with the Space Telescope Imaging Spectrograph on the \textit{Hubble Space Telescope}.
We performed a cross-match with a radius of $1.0''$, identifying \textit{Gaia} DR3 counterparts for all 21 O-type stars. Among them, for the three stars located at the innermost center, only two stars (\texttt{SSN22} and \texttt{SSN43}), excluding the brightest star (\texttt{SSN14}), do not have available $G_\mathrm{BP}-G_\mathrm{RP}$ values.
These three stars are clustered within a radius of $1.3''$, which is smaller than the $2.1''$ window size of the $G_\mathrm{BP}$ and $G_\mathrm{RP}$ measurements. 
\texttt{SSN22} and \texttt{SSN43} can be interpreted as examples where the completeness of the $G_\mathrm{BP}$ and $G_\mathrm{RP}$ measurements is compromised due to crowding.
The proportion of stars for which $G_\mathrm{BP}-G_\mathrm{RP}$ is unavailable due to crowding is less than 10\%, even in the central region of NGC~346, the most crowded area in the SMC.
Thus, unavailable $G_\mathrm{BP}-G_\mathrm{RP}$ does not result in a significant decrease in completeness for the selection across the entire SMC.
We present a proportion of O-type stars identified by \citet{2022A&A...666A.189R} that are included in our massive star catalog in \secref{sec:3.2.5}.

\begin{figure*}
 \centering
 \includegraphics[width=18cm,clip]{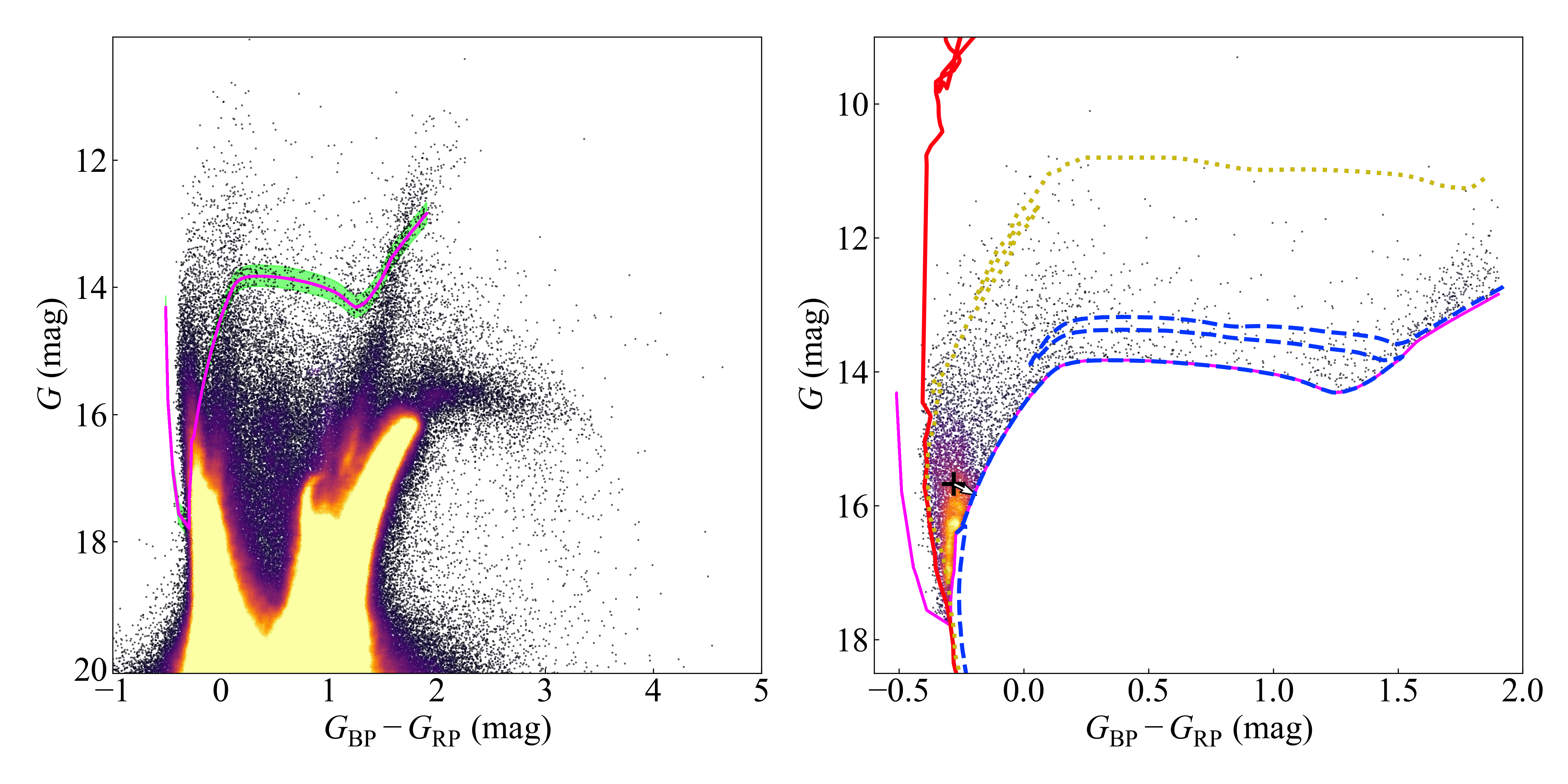}
 \caption{
 ($G_\mathrm{BP}-G_\mathrm{RP}$, $G$) CMD of the SMC region excluding foreground sources.
 The area above the magenta line is the region occupied by massive stars in the CMD, defined in the \secref{sec:data1}.
 In the left panel, stars in the SMC region are plotted excluding foreground objects.
 The green shadow represents the range of variation in the massive star selection region for a distance modulus of $\pm5~\mathrm{kpc}$, which corresponds to distances of 60~kpc and 70~kpc.
 In the right panel, only massive star candidates are plotted.
 The red line, yellow dotted line, and blue dashed line are PARSEC isochrones for ages of $1.0$, $10.0$, and $36.5~\mathrm{Myr}$, respectively, with a reddening correction of $A_{V}=0.15~\mathrm{mag}$ added, as noted in the text.
 The black cross represents the median value $(-0.2809, 15.67)$ of the massive star candidates.
 The white arrow extending from the black cross is the reddening vector corresponding to $A_{V}=0.15~\mathrm{mag}$, illustrating the amount of reddening required to move from the median position $(-0.2809, 15.67)$ across the magenta line.
}
\label{CMD1}
\end{figure*}

\section{\textit{Gaia} selected massive star candidates}
\label{sec:selection}
In this section, we discuss the selection of massive stars using the \textit{Gaia} CMD.
\secref{sec:CMD} describes the selection method, and \secref{sec:CMD_result} shows the results of this selection.
In \secref{sec:prop}, we outline the properties of massive star candidates expected from the selection method.
As preparatory work for scientific discussions involving these candidates, \secref{sec:spacial} and \secref{sec:bonanos} demonstrate the validity of the selection method discussed in \secref{sec:CMD}.

\subsection{Massive Star Selection Using CMD}
\label{sec:CMD}

In this section, we describe the method for selecting massive stars in the SMC using the $G$ versus $G_\mathrm{BP}-G_\mathrm{RP}$ CMD based on data from \textit{Gaia} DR3. In the subsequent discussions, we only use 1,832,516 objects for which data on $G$, $G_\mathrm{BP}$ and $G_\mathrm{RP}$ are available.

The left panel of \figref{CMD1} shows ($G_\mathrm{BP}-G_\mathrm{RP}$, $G$) CMD of the SMC region, with foreground objects removed.
We will define the region on this CMD that corresponds to massive stars with initial masses $M_\mathrm{ini}\geq8M_{\odot}$ and identify the stars within this region as massive star candidates.
The advantage of using optical ($G_\mathrm{BP}-G_\mathrm{RP}$, $G$) CMD for massive star selection is not only the high photometric accuracy of \textit{Gaia}, but also the fact that optical CMD is more sensitive to the selection of massive stars than near-infrared; the peak of the blackbody of massive stars is at shorter wavelengths and the optical CMD is more sensitive to color changes of massive stars.
Conversely, optical observations are more affected by reddening than near-infrared observations. However, particularly in the SMC, where the metallicity is low, the areas significantly affected by dust-induced reddening occupy a relatively small fraction of the overall region, as will be assessed later. Therefore, most massive stars in regions except for a limited number of star-forming regions are expected to be detectable using this method.

The definition of the region on the CMD where massive stars are located is based on the theoretical models of stellar evolutionary tracks from PARSEC (version 1.2S; \citealp{2012MNRAS.427..127B,2014MNRAS.444.2525C,2015MNRAS.452.1068C, 2014MNRAS.445.4287T, 2017ApJ...835...77M, 2019MNRAS.485.5666P, 2020MNRAS.498.3283P}).
PARSEC calculates evolutionary tracks for stars based on their initial masses by specifying input parameters such as the initial mass function (IMF) and metallicity, allowing for the derivation of isochrones\footnote{The models calculated by PARSEC can be downloaded from (\href{http://stev.oapd.inaf.it/cgi-bin/cmd}{http://stev.oapd.inaf.it/cgi-bin/cmd}) by specifying various parameters.}.
The PARSEC outputs contain information on the $M_\mathrm{ini}$ of each star that forms the isochrones. We use this $M_\mathrm{ini}$ defined by PARSEC to determine the region on the CMD occupied by stars with masses greater than 8$M_\mathrm{\odot}$.
We specified the input parameters for PARSEC corresponding to the stars in the SMC as the IMF from \citet{2001MNRAS.322..231K, 2002Sci...295...82K}, and a metallicity of $Z=0.1Z_{\odot}$, where $Z_{\odot}=0.0152$.
$Z=0.1 Z_{\odot}$ represents the average metallicity of the SMC, dominated by older stars (e.g., \citealp{2014MNRAS.442.1680D, 2020MNRAS.497.3746C}), and corresponds to the average metallicity of young stars in the wing structure \citep{2018MNRAS.478.5017R}. As metallicity increases, isochrones tend to shift redward; by using a lower metallicity to define the region of massive stars on the CMD, we can avoid contamination from blue, lower metallicity intermediate- to low-mass stars.
The ages of the stars in the PARSEC tracks were outputted in the range of $Age = 0.1 \text{--} 50.0~\mathrm{Myr}$ at intervals of $0.1~\mathrm{Myr}$. This range of $Age$ corresponds to $logAge \simeq 5.0 \text{--} 7.7$.

\begin{figure*}
 \centering
 \includegraphics[width=18cm,clip]{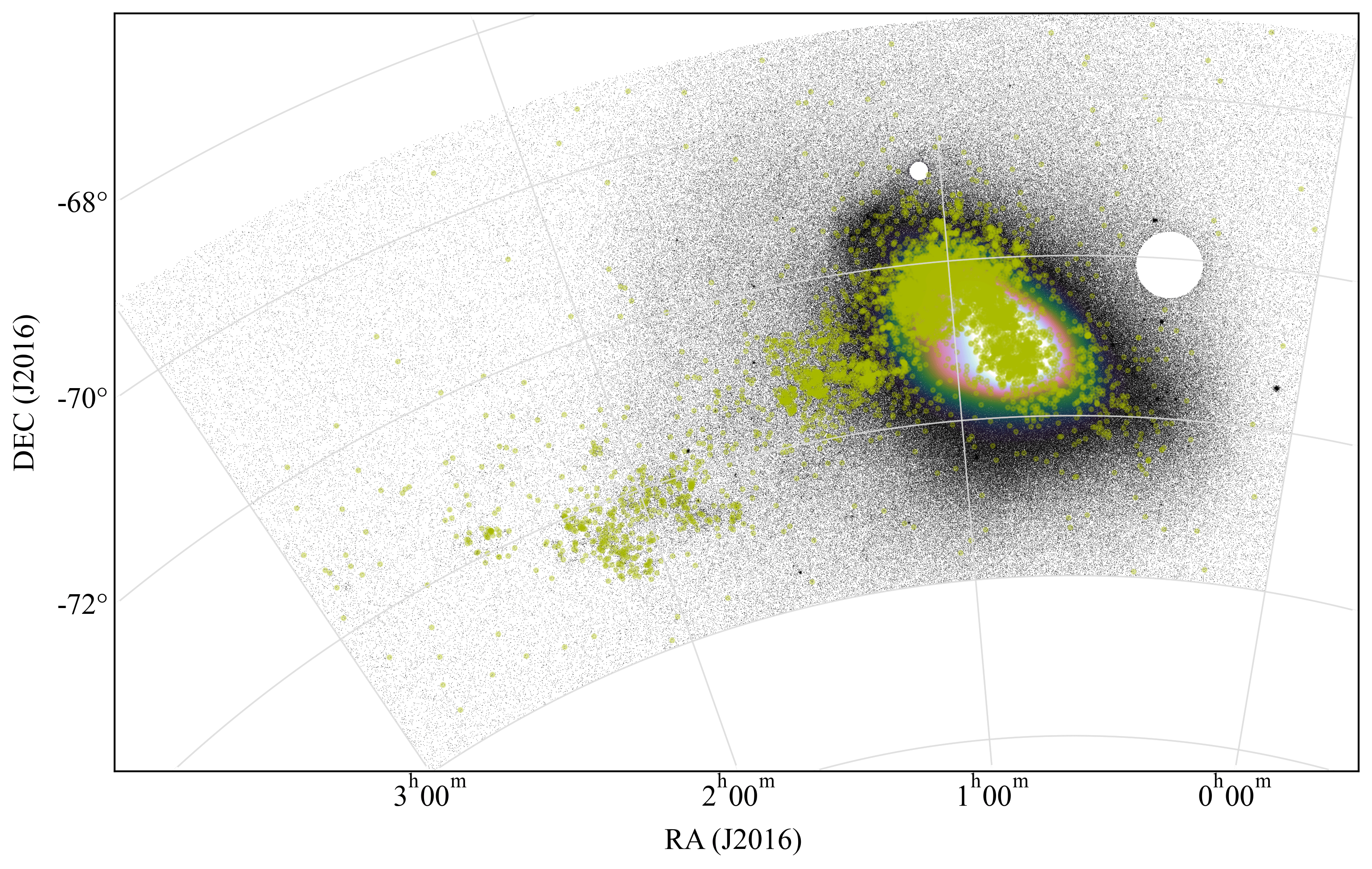}
 \caption{
 The spatial distribution of the massive star candidates selected in \secref{sec:data1}. On the background of \figref{SMC_Gaia}, the massive star candidates are overplotted as dark yellow circles.
}
\label{SMC_Gaia_withOB}
\end{figure*}

The region of the CMD where massive stars are located is defined by two contours: the border on the bluer side, determined by unreddened O-type main sequence stars, and the border that separates intermediate- to low-mass stars from massive stars, which is influenced by reddening from foreground dust in the Milky Way.
The border for the former is derived by selecting stars from the PARSEC outputs with $Age = 0.1~\mathrm{Myr}$ and $18 M_{\odot} \leq M_\mathrm{ini} \leq 60 M_{\odot}$ as O-type main sequence stars, which were then interpolated on the CMD. Massive stars in the SMC are expected to be affected by reddening from Milky Way foreground dust and SMC dust, making the unreddened O-type main sequence stars function as the bluest boundary.
To determine the more critical latter border, we added a reddening correction of $A_{V} = 0.15~\mathrm{mag}$ (e.g., \citealp{2022MNRAS.511.1317C}), attributed to the Milky Way foreground, to the outputted PARSEC tracks using \texttt{dustapprox} \citep{Fouesneau_dustapprox_2022} \footnote{The reason for not applying individual reddening corrections based on the reddening map is that the reddening map of the SMC, which extends along the line-of-sight, can be biased by the reddening of foreground stars, making it less likely to reflect the reddening of structures deeper within the galaxy \citep{2022MNRAS.511.1317C}. Additionally, young stars are often surrounded by localized ISM, and using reddening values calculated from the average of surrounding stars could compromise the homogeneity of the selection.
}.
\texttt{Dustapprox} is capable of calculating the integrated extinction effect based on the estimated stellar parameters outputted by PARSEC and the shape of the $G$, $G_\mathrm{BP}$, and $G_\mathrm{RP}$ passbands. We utilized the precomputed \textit{Gaia} EDR3 passbands model provided by \texttt{dustapprox}. This model assumes the extinction law from \citet{1999PASP..111...63F} and an $R_{V} = 3.1$, which are used for the Milky Way not the SMC. However, we emphasize that the assumptions regarding the extinction law and $R_{V}$ do not significantly impact the results. This is because the wavelengths of the \textit{Gaia} passbands (even the short-wavelength $G_\mathrm{BP}$ band, around $\sim$$500~\mathrm{nm}$) are sufficiently long compared to the peak at 217.5~nm (which depends on the dust composition) in the extinction curve, resulting in minimal uncertainty from the extinction law. For the same reason, the uncertainty introduced by $R_{V}$ is also small; the change in $A_\mathrm{500\,nm}/A_{V}$ when varying $R_{V}$ from 3.0 to 2.0 under the extinction law of \citet{1999PASP..111...63F} is approximately 5\%, translating to an uncertainty of only 0.01~mag for $A_{V} = 0.15~\mathrm{mag}$.
Then, we removed stars with $T_\mathrm{eff} \geq 50000~\mathrm{K}$ from the PARSEC outputs, as these stars are not included in the atmosphere models used by \texttt{dustapprox} \citep{2003IAUS..210P.A20C}.
From the reddening-corrected PARSEC outputs described above, we extracted the border where there is no contamination from intermediate- to low-mass stars with $M_\mathrm{ini} < 8 M_{\odot}$. 
We applied a cutoff at $G_\mathrm{BP}-G_\mathrm{RP} \gtrsim 1.9~\mathrm{mag}$ due to the discontinuity of the PARSEC tracks in this region, as well as the anticipated contamination from evolved AGB stars, which are expected to represent intermediate- to low-mass stars.
The combination of the two defined borders, along with a distance modulus of 19.06~mag corresponding to 65~kpc, is represented by the magenta line overplotted on the CMD in \figref{CMD1}. Utilizing the distance modulus for 65~kpc allows us to detect slightly fainter massive stars located beyond the main body, at $\mathrm{\sim}60~\mathrm{kpc}$.

In the CMD of \figref{CMD1}, we selected 7,426 stars as massive star candidates located in the region above the magenta line.
Note that, the number of selected massive star candidates is 6,481 for a distance modulus of 18.89~mag (corresponding to 60~kpc) and 8,418 for a distance modulus of 19.23~mag (corresponding to 70~kpc). The $\pm5~\mathrm{kpc}$ variation in distance modulus changes the number of massive star candidates by approximately $\pm1,000$, having little impact on the majority of the sample. The variation in the selection region due to changes in the distance modulus of $\pm5~\mathrm{kpc}$ is shown by the green shadow in the left panel of \figref{CMD1}.

To prevent contamination from intermediate- to low-mass stars with significant photometric errors, we did not take into account the error bars in the $G$ and $G_\mathrm{BP}-G_\mathrm{RP}$.
The number of objects with errors exceeding either of the typical errors mentioned in \secref{sec:data1}, $\sigma_{G} \approx 0.0035~\mathrm{mag}$ and $\sigma_{G_\mathrm{BP}-G_\mathrm{RP}} \approx 0.0155~\mathrm{mag}$, is 640 (less than 9\% of the total), indicating that no significant bias due to photometric errors exists in the selected sample.

\renewcommand{\arraystretch}{1.2}
\begin{deluxetable*}{lcccccccccccc}
 \tablecaption{Catalog of 7,426 Massive Star Candidates in the SMC}
 \tablewidth{0pt}
 \label{catalog}
\tablehead{
    \shortstack{} &
    \shortstack{Source ID} &
    \shortstack{RA (J2000)} &
    \shortstack{DEC (J2000)} &
    \shortstack{$G$} &
    \shortstack{$\sigma_{G}$} &
    \shortstack{$G_\mathrm{BP}$} &
    \shortstack{...} &
    \shortstack{RV} &
    \shortstack{$\sigma_\mathrm{RV}$} &
    \shortstack{...} &\\
  &
  &
 \colhead{(deg)}&
 \colhead{(deg)}&
 \colhead{(mag)}&
 \colhead{(mag)}&
 \colhead{(mag)}&
 \colhead{ }&
 \colhead{($\mathrm{km~s}^{-1}$)}&
 \colhead{($\mathrm{km~s}^{-1}$)}&
 \colhead{ }&
}
\startdata
0 & 4684253618358645888 & 6.85023967649 & -75.92770891751 & 13.176072 & 0.002765 & 13.889589 & ... & 153.98 & 0.98  & ...\\
1 & 4637126484111349632 & 30.96574913575 & -75.90096767348 & 13.021945 & 0.002766 & 13.561051 & ... & 275.50 & 1.73 & ...\\
2 & 4684982216610572928 & 3.22812911558 & -75.83310782468 & 13.773994 & 0.002761 & 14.299227 & ... & 321.78 & 3.66 & ...\\
3 & 4637745788332390528 & 23.29629839582 & -75.70372375109 & 14.094418 & 0.002762 & 14.656014 & ... & 69.80 & 2.78 & ...\\
\enddata
\tablenotetext{ }{Here we only show part of the table. The machine readable table will be available in the HTML version of the article.}
\tablecomments{The complete table includes the following: the \textsc{source\_id} of \textit{Gaia} DR3 (\texttt{Source-ID}), coordinates in J2000 (\texttt{RAdeg}, \texttt{DEdeg}), photometry for $G$, $G_\mathrm{BP}$, and $G_\mathrm{RP}$ (\texttt{Gmag}, \texttt{GBPmag}, and \texttt{GRPmag}), the parallax and its uncertainty (\texttt{plx}), the \textit{Gaia} PMs (\texttt{pmRA}, \texttt{pmDE}), flags of the PM 2$\sigma$-clipped samples (\texttt{PM-2sigma-clipped}; see \secref{sec:pm}), the internal PMs within the SMC calculated in \secref{sec:pm} (\texttt{pmRA-in}, \texttt{pmDE-in}), \textit{Gaia} radial velocities and their uncertainties (\texttt{RVel}; before correction for hot stars; see \secref{sec:RV_OB}), the supercluster labels (\texttt{Supercluster-label}; see \secref{sec:identification_st}), the information on the success or failure of the cross-match between \citet{2010AJ....140..416B} and \textit{Gaia} DR3 for cross-match radii of $1.0''$ and $0.5''$ (\texttt{B10-match-1s} and \texttt{B10-match-05s}), and the names and spectral type classifications of the objects listed in \citet{2010AJ....140..416B} for successfully cross-matched objects (\texttt{B10-Name} and \texttt{B10-CC}). When two objects listed in \citet{2010AJ....140..416B} are cross-matched with a single massive star candidate, their names and spectral type classifications are distinguished by ``!!". The duplication of two objects in the cross-match using a radius of $0.5''$ arises from the issue that the same object is listed as separate entities in \citet{2010AJ....140..416B}.}
\end{deluxetable*}
\renewcommand{\arraystretch}{1.0}

\subsection{Result of Massive Star Candidate Selection}
\label{sec:CMD_result}
This section presents the massive star catalog obtained through the selection method discussed in \secref{sec:CMD}, along with some analyses that can be immediately derived from the catalog.

\subsubsection{Spatial Distribution of Massive Star Candidates}
\label{sec:3.2.1}
\figref{SMC_Gaia_withOB} shows the spatial distribution of massive star candidates in the SMC region. A catalog of the 7,426 massive star candidates is provided in \tabref{catalog}.
Our massive star candidates exhibit a spatial distribution distinct from that of the old stellar population, which traces the bar structure and wing structure of the ISM, as discussed in \secref{sec:spacial}.
Additionally, many massive star candidates are distributed along the Magellanic Bridge.
At $\mathrm{RA}=02\mathrm{h}\textsc{--}03\mathrm{h}$ along the bridge, a collective distribution of massive star candidates corresponds to the positions of three tidal dwarf galaxies identified by \citet{1995ApJS..101...41B, 2020AJ....159...82B}, which are thought to have formed from tidal debris of the SMC. 
However, these may include some intermediate-mass stars ($\lesssim$$8M_\odot$), as the stars in the bridge are positioned closer to us than those in the main body of the SMC; closer stars appear brighter, making it impossible to separate distant massive stars from nearby intermediate-mass stars on the CMD.

\subsubsection{Impact of Blending}
\label{sec:3.2.2}
As explained in \secref{sec:data2}, blending may affect the measurements of $G_\mathrm{BP}$ and $G_\mathrm{RP}$ in crowded regions.
Among the 7,426 massive star candidates, 121 objects have a photometric excess factor $C>1.4$.
This accounts for less than 2\%, and thus does not significantly affect the overall properties.
In parallel, there are four pairs of massive star candidates that are closer than $1.75''$, and could be blended depending on the scan direction. Among these, three pairs are always blended, as they are closer than $1.05''$.
If false detections of intermediate- to low-mass stars due to blending occur in our selection, these stars would likely be located near blue massive stars. Therefore, the number of false detections of intermediate- to low-mass stars due to blending is at most around four.
Note that, using the fitting results of $C$ for isolated stars from \citet{2021A&A...649A...3R}, the average $C$ expected for massive star candidates with $G_\mathrm{BP}-G_\mathrm{RP}<0.5$, asssuming no blending is approximately 1.15. Therefore, $C = 1.4$ corresponds to an excess of about 20\% in the $G_\mathrm{BP}$ and $G_\mathrm{RP}$ fluxes. A 20\% increase in flux corresponds to a brightening of approximately 0.20~mag, resulting in a maximum change of 0.20~mag in $G_\mathrm{BP}-G_\mathrm{RP}$.

\subsubsection{Validity of Globular Cluster Cutout}
\label{sec:3.2.3}
There are 325 stars within the region cut out as a globular cluster that are included in the massive star selection area based on the CMD. 
Among these, 323 stars are collectively distributed on the red side of the CMD, with $G_\mathrm{BP}-G_\mathrm{RP} \gtrsim 0.89~\mathrm{mag}$, indicating that they are comprised of evolved stars. The remaining two stars have $G_\mathrm{BP}-G_\mathrm{RP} = 0.578$ and $0.548~\mathrm{mag}$, which indicate that they are evolving into yellow giants. These stars are located at the center of NGC~104, consistent with the concentration of yellow supergiants due to mass segregation in the globular cluster center.
Furthermore, all 325 stars within the globular cluster region have proper motions that differ in direction from the average proper motion of other massive star candidates (which is close to that of the SMC) by more than 2$\sigma$ (see \secref{sec:pm}). This suggests that it is more consistent to consider them as belonging to the foreground globular cluster rather than the SMC.

\subsubsection{Detection Rate of Bright Far-Ultraviolet Sources}
\label{sec:3.2.4}
Our massive star catalog includes $\approx$85\% of the photometrically identified massive stars in the far-ultraviolet (FUV) provided by \citet{2024AJ....168..255H}.
\citet{2024AJ....168..255H} provided a FUV catalog of 76,838 objects in the SMC using the Ultra Violet Imaging Telescope onboard AstroSat. Among these, stars brighter than 15~mag in FUV luminosity were interpreted as massive stars with a mass greater than 8$M_\odot$.
Since we were unable to find many objects in the SMC region that matched the \texttt{Gaia\_DR3\_Source\_ID} from \citet{2024AJ....168..255H}, we independently performed a cross-match within a radius of $1.0''$ (the same radius used by \citealt{2024AJ....168..255H}) and identified 84,566 objects in the \textit{Gaia} DR3.
Here, a single FUV source may correspond to multiple \textit{Gaia} DR3 objects.
Among the 84,566 objects, 2,180 are identified by \citet{2024AJ....168..255H} as stars in the SMC with FUV luminosities brighter than 15~mag. Of these, 2,114 have $\varpi < 0.1~\mathrm{mas}$ and usable $G_\mathrm{BP}-G_\mathrm{RP}$ from \textit{Gaia} DR3 data.
Our massive star candidates account for 1,842 out of the 2,114 objects. The $\approx$300 undetected objects are localized on the CMD around $G \simeq 14$--$16~\mathrm{mag}$ and $G_\mathrm{BP}-G_\mathrm{RP} \simeq -0.2$ to $0.2~\mathrm{mag}$, suggesting that our method using the CMD imposes more stringent criteria.

\subsubsection{Detection Rate of O-type Stars in the Most Crowded Regions}
\label{sec:3.2.5}
Our massive star catalog includes 16 of the 21 spectroscopically identified O-type stars located in the core of NGC~346, which were reported by \citet{2022A&A...666A.189R}, with the details of these stars described in \secref{sec:data2}.
The detection rate for the 19 stars with available $G_\mathrm{BP}-G_\mathrm{RP}$ is approximately 84\%. 
Of the three O-type stars that are not detected, \texttt{SSN14} is located within a radius of $1.3''$, along with two other stars for which $G_\mathrm{BP}-G_\mathrm{RP}$ are unavailable.
Due to its large parallax ($\varpi\approx1.2$), \texttt{SSN14} is excluded as a foreground object.
\texttt{SSN14} may have had its parallax measurement hindered by dynamical interactions with nearby objects.
The other two undetected O-type stars are plotted just outside the red edge of the massive star detection region in the CMD, where significant reddening in the central star formation region may restrict their detection.
A comparison with \citet{2010AJ....140..416B}, which provides the most comprehensive spectroscopic catalog of massive star in the SMC, is described in \secref{sec:bonanos}.

\begin{figure*}
 \centering
 \includegraphics[width=18cm,clip]{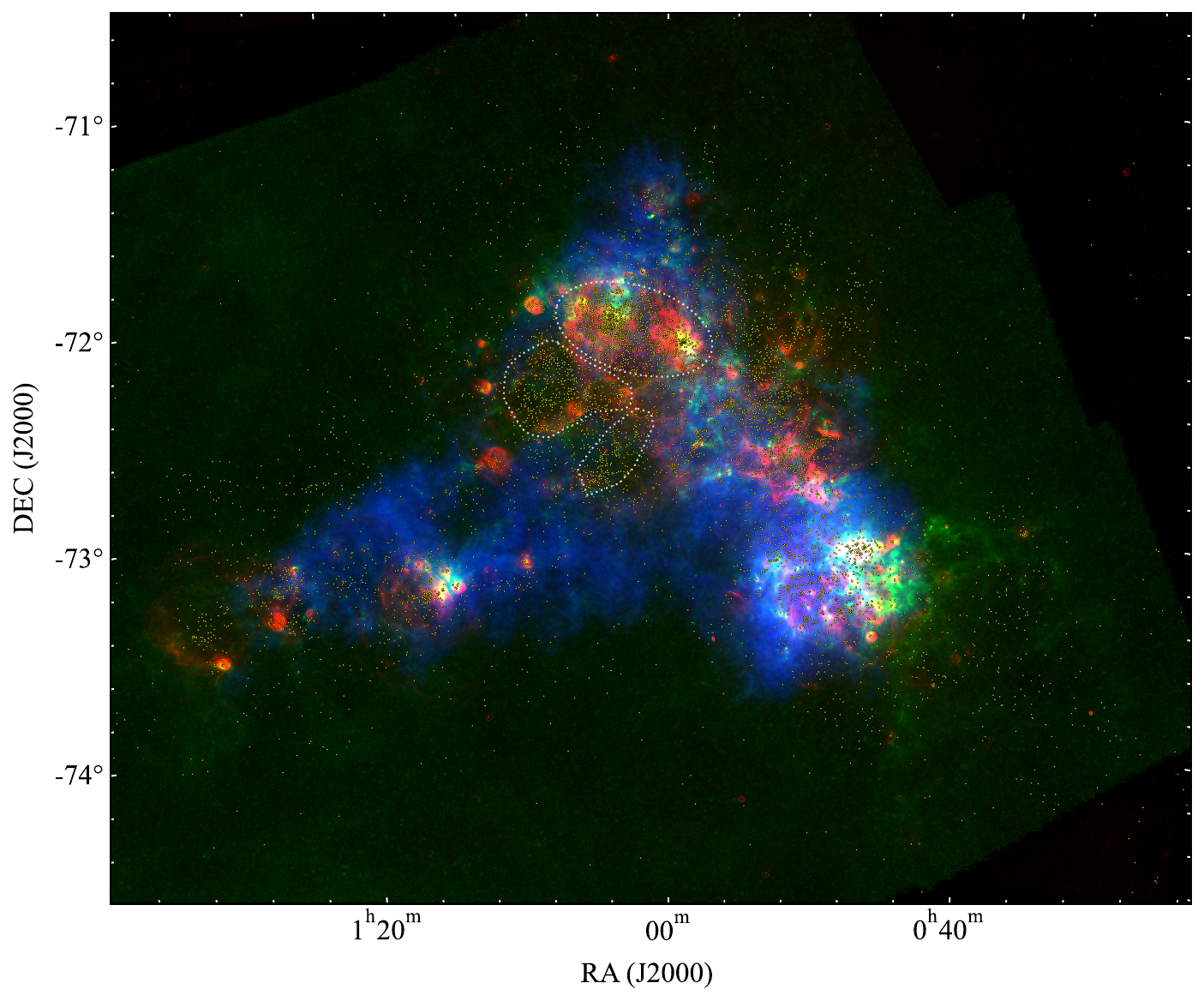}
 \caption{
 Comparison of the spatial distribution of massive star candidates with that of the ISM around the SMC.
 The background is a color multiwavelength montage of the H$\alpha$ emission by MCELS (red; 0--$300\times10^{-17}~\mathrm{ergs~cm^{-2}~s^{-1}}$; \citealp{1999IAUS..190...28S}) with the \textit{Herschel} 350~\textmu m image (green; 0--25~$\mathrm{MJy~sr}^{-1}$; \citealp{2013AJ....146...62M}) and the H\,\textsc{i} with ASKAP \& Parkes integrated over the LSRK velocity range 130--170~$\mathrm{km~s^{-1}}$ (blue; 1,000--3,000~$\mathrm{K~km~s}^{-1}$; \citealp{2018NatAs...2..901M}). The resolution varies by color. Massive star candidates are overplotted as yellow circles. The three regions indicated by white dotted lines correspond to the three overdensities in the northern part, which are highlighted in \figref{KDE}.}
\label{3color}
\end{figure*}

\subsection{Overall Properties of Selected Massive Star Candidates}
\label{sec:prop}

In this section, we present the approximate ranges of $A_{V}$ and $Age$ for the massive star candidates.
The right panel of \figref{CMD1} shows a CMD plotted exclusively for the massive star candidates.
The median value of the massive star candidates shown in the right panel of \figref{CMD1} is $(-0.2809,\,15.67)$. When applying an additional reddening correction of $A_{V} = 0.15~\mathrm{mag}$, alongside the Milky Way foreground reddening of $A_{V} = 0.15~\mathrm{mag}$ already accounted for, the resulting value falls outside the magenta selection area. Therefore, at first glance, the reddening coverage for this massive star selection appears to be $A_{V} \lesssim 0.30~\mathrm{mag}$.
However, we note that it is possible to select massive stars that have experienced greater reddening.
This is because the majority of the massive star candidates used to calculate the median are expected to be affected by reddening originating from within the SMC. The median $(-0.2809,\,15.67)$ corresponds to a main-sequence star with an initial mass of approximately $M_\mathrm{ini}\simeq28M_{\odot}$, which is expected to have experienced a total reddening of about $A_{V}\simeq0.33$~mag, including contributions from within the SMC.
This implies that for a main-sequence star with an initial mass of approximately $M_\mathrm{ini}\simeq28M_{\odot}$, a reddening of about $A_{V} \simeq 0.48$~mag would be required to shift it out of the magenta selection region. 

According to \citet{2022MNRAS.511.1317C}, even in regions of significant reddening within the SMC, $A_{V}$ is typically less than or equal to 0.3~mag ($E(B-V)\lesssim0.1~\mathrm{mag}$; in the case of assuming $R_{V}=3.1$), with areas exceeding $A_{V} = 0.48$~mag being found only in limited star formation regions in the northern and southwestern parts of the bar.
One of the regions with the highest $A_{V}$ is the NGC~346 region, where \citet{2022A&A...666A.189R} performed spectral energy distribution fitting for 19 O-type stars located in its central area. 
The fitting results indicate a mean reddening of $A_{V} = 0.40~\mathrm{mag}$ ($E(B-V) = 0.13$), with two stars exceeding $A_{V} = 0.48~\mathrm{mag}$.
Note that the two stars listed by \citet{2022A&A...666A.189R} that are not included in our catalog, as described in \secref{sec:3.2.5}, have $A_{V} = 0.372~\mathrm{mag}$ and are the faintest O-type stars (spectral types O9V and O9.5V).
Additionally, several studies suggest that the average $A_{V}$ in the SMC, including for various stellar populations containing massive stars, is 0.3~mag or smaller (e.g., \citealp{1995ApJ...438..188M, 2012ApJ...744..128S, 2018MNRAS.473.3131M, 2020ApJ...889..179G, 2021A&A...646A.106S}; $E(B-V)\lesssim0.1~\mathrm{mag}$).
Therefore, the number of massive star candidates missed in this selection due to reddening is expected to be limited. Moreover, the notion that the median, being redder than the main sequence, is influenced by internal reddening in the SMC is consistent with the fact that stars spend the majority of their lives on the main sequence. 
However, even taking into account that the median is somewhat influenced by bright red supergiants, it is important to recognize that $28M_{\odot}$ is too heavy to be considered a representative $M_\mathrm{ini}$ for the massive star candidates.

The right panel of \figref{CMD1} shows the overplotted PARSEC isochrones for ages of $1.0$, $10.0$, and $36.5$~Myr, with a reddening correction of $A_{V}=0.15$~mag added. The PARSEC isochrone for $Age=1.0$~Myr aligns with the blue edge of the distribution of massive star candidates in the CMD, indicating that the reddening due to the Milky Way's foreground, $A_{V}=0.15$~mag, affects the entire SMC. The PARSEC isochrone for $Age=36.5$~Myr constitutes a significant portion of the magenta border, suggesting that the ages of the selected massive star candidates are $Age\lesssim\mathrm{36.5~Myr}$. There is also a possibility that a small number of massive stars with ages less than 50.0~Myr, located in the blue loop, are included.
As the age increases, even smaller amounts of reddening can result in exclusion from the current selection of massive stars.

\begin{figure*}
 \centering
 \includegraphics[width=12cm,clip]{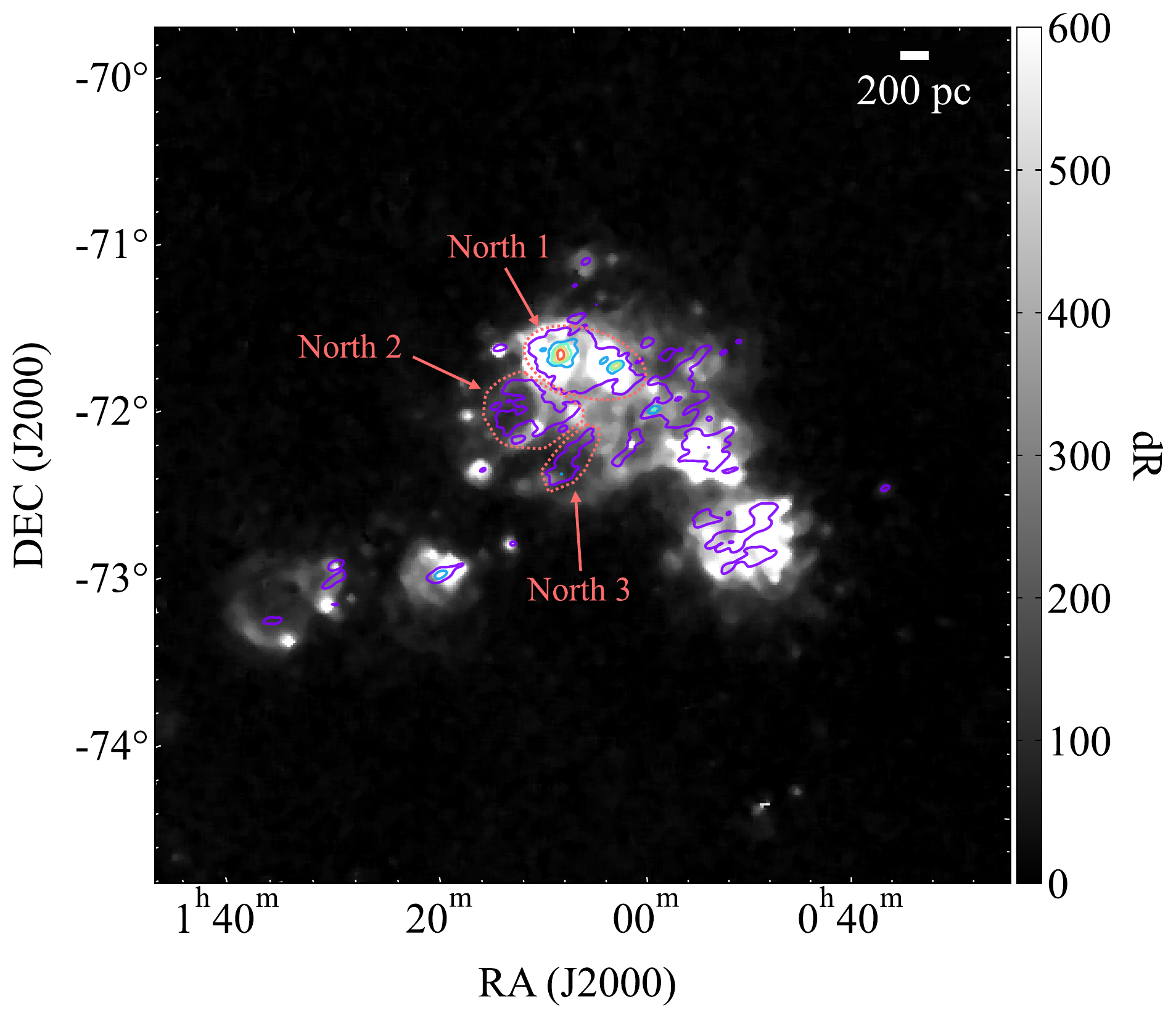}
 \caption{
 KDE map of our massive star candidates. The contours show [0.01, 0.1, 0.3, 0.5, 0.7, 0.9, 1.0] levels, normalized to the maximum value. 
 The background shows H$\alpha$ emission by SHASSA in the range of 0--600~deciRayleighs (dR; \citealp{2001PASP..113.1326G}). Three distinct peaks located at the boundaries of the bar structure and wing structure in the northern region are highlighted with dashed lines. Note that the contours are elongated in the east-west direction due to the influence of coordinate distortion.
}
\label{KDE}
\end{figure*}

\begin{figure*}
 \centering
 \includegraphics[width=18cm,clip]{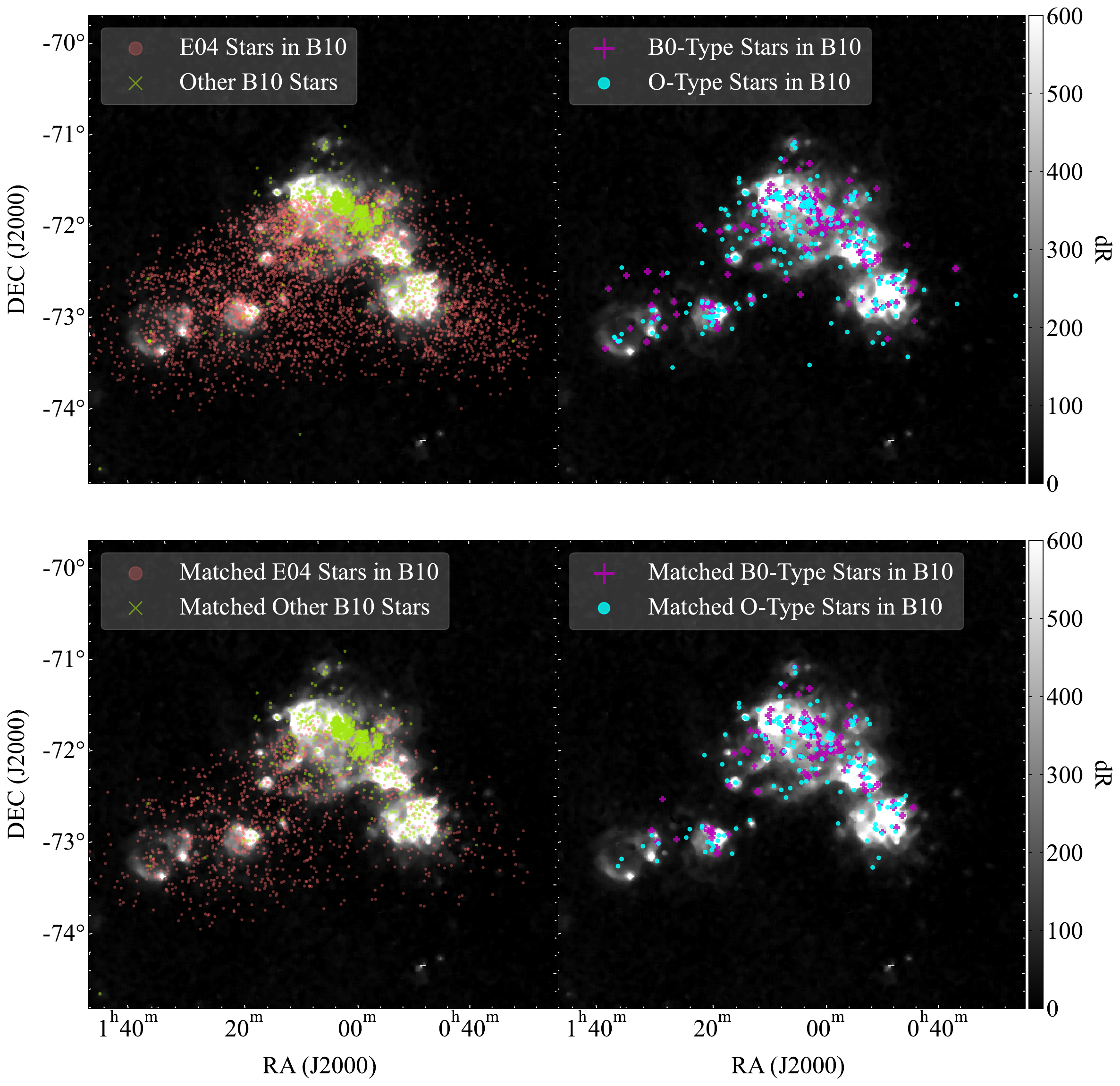}
 \caption{
  (Upper left) The spatial distribution of spectroscopically identified massive stars around the SMC from the B10 catalog. The background shows 0--600~dR of H$\alpha$ emission by SHASSA. E04 stars, which are part of the 4,073 objects in the B10 catalog, are represented by red circles, while the remaining 1,251 objects are represented by green crosses.
  (Upper right) The spatial distribution of stars classified as B0-type and O-type from the B10 catalog around the SMC. The background is the same as in the upper left panel. Stars classified as B0-type (181 objects) are represented by purple crosses, and O-type stars (281 objects) are represented by light blue circles. 
  Since there is a range of type classifications, stars that may be later than B3, including B3 itself, are excluded. Stars that are classified as both O-type and B0-type are included in the O-type category.
  (Lower left) Similar to the upper left panel, but showing B10 catalog stars successfully cross-matched in \textit{Gaia} DR3 data. Among the 4,073 E04 stars, 1,323 successfully cross-matched stars are represented by red circles, while among the 1,251 stars identified through other observations, 1,142 successfully cross-matched stars are represented by green crosses.
  (Lower right) Similar to the upper right panel, but showing B10 catalog stars successfully cross-matched in \textit{Gaia} DR3 data. The purple crosses represent 130 cross-matched B0-type stars, and the light blue circles represent 188 cross-matched O-type stars.
}
\label{B10_E04}
\end{figure*}

\subsection{Comparison of the Spatial Distribution of the ISM and Massive Star Candidates}
\label{sec:spacial}
In this section, we detail the spatial distribution of the selected massive star candidates and compare it with that of the ISM, thereby validating the selection method.
\figref{3color} is an enlarged view of the spatial distribution of massive star candidates around the SMC, compared to the ISM (emission of H$\alpha$, dust, and H\,\textsc{i}; integrated over an H\,\textsc{i} velocity range centered on a systemic velocity of $150~\mathrm{km~s^{-1}}$ with a width of $\pm20~\mathrm{km~s^{-1}}$, referenced to the LSRK). Our massive star candidates exhibit a spatial distribution that aligns with the ISM, tracing the bar and wing structures of the SMC. This alignment is also clearly discernible from the differences in the distribution of all stars in the SMC, as shown in \figref{SMC_Gaia_withOB}.
The match with the distribution of the ISM is a characteristic expected for young massive stars, demonstrating the validity of our selection.

\figref{KDE} illustrates the surface density map of our massive star candidates, derived from kernel density estimation (KDE), with the H$\alpha$ image as the background.
In constructing \figref{KDE}, we note that a bandwidth of 30~pc was employed for the Gaussian kernel to emphasize the distribution's alignment with the H$\alpha$ emission.
Notably, not only is there a general agreement in the overall distribution, but the peaks in surface density of our massive star candidates align with several local maxima of H$\alpha$ emission.
This includes the H$\alpha$ brightest regions in the northern and southwestern areas, as well as the outer edge regions of the SMC, such as the wing structure.
These correspond to star clusters that contain many massive stars.
Since H$\alpha$ emission arises from the radiation of H\,\textsc{ii} regions formed by massive stars, the spatial distribution of our massive star candidates matching the H$\alpha$ emission supports the validity of our selection method.

The region marked as North~1 in \figref{KDE} is where the brightest H\,\textsc{ii} region in the SMC is located, and it is also the area with the highest density of our massive star candidates. Within North~1, two prominent peaks can be observed, corresponding to well-known massive star clusters (NGC~371 to the east and NGC~346 to the west). The northern region of the SMC is currently the most active star formation region \citep{2018MNRAS.478.5017R}, making it reasonable to find a significant number of massive star candidates in this region.
The regions marked as North~2 and North~3 are located slightly towards the wing structure from North~1, exhibiting a clear density gap. Additionally, the emission of H$\alpha$, 350~\textmu m and H\,\textsc{i} in the North~2 and North~3 regions are weak and appear to exhibit a shell-like structure (see \figref{3color}). 
Especially, the H$\alpha$ along the southern boundary of North~2 and the H\,\textsc{i} along the northern boundary emphasize its spherical shape.
The excess density in the North~2 and North~3 regions is also observed in the KDE map of VISTA near-infrared selected upper main-sequence stars \citep{2018ApJ...858...31S} and in some \textit{Gaia} selected young populations \citep{2021A&A...649A...7G}. However, we found very few descriptions focusing on the North~2 and North~3 regions in previous studies.
\citet{2001A&A...379..864M} and \citet{2024AJ....168..255H} refer to the North~2 and North~3 regions collectively as a ``shell,'' but they do not discuss the separation between North~2 and North~3, nor do they compare these regions with the ISM. 
A detailed discussion of the North~2 and North~3 regions will be revisited in \secref{sec:dbscan}.

In specific regions, as clearly seen in the southwestern region, our massive star candidates tend to be distributed in a way that avoids the green 350~\textmu m dust emission. In \figref{3color}, the absence of massive star candidates highlights the filamentary structures in the southwest and the clump structures in the north, which are prominently displayed in green. The emphasis on green in the three-color map also indicates that the red H$\alpha$ emission is weak, suggesting that regions exhibiting an anti-correlation between our massive star candidates and dust emission are likely to have large optical thickness. Specifically, the presence of massive stars may be obscured by dust, or it is possible that no massive stars have yet formed due to their relatively young evolutionary stage. This anti-correlation between our massive star candidates and dust emission implies, as demonstrated in \secref{sec:prop}, that our selection method is not applicable under conditions of substantial reddening ($A_{V}\geq0.5~\mathrm{mag}$). The northern and southwestern regions correspond to the peaks in the reddening map from \citet{2022MNRAS.511.1317C}. Areas of large reddening fall outside the scope of this optical study. The incompleteness of our selection due to reddening will be discussed in the following subsection.

We note that the KDE map of VISTA near-infrared selected upper main-sequence stars, created using a bandwidth of 10~pc from \citet{2018ApJ...858...31S}, shows a localized excess density of young stars even in regions of large reddening in the southwestern area. This may indicate that the near-infrared selection compensates for the incompleteness of our method; however, several caveats must be considered when comparing it to our \figref{KDE}.
Firstly, there is a difference in the sample. While \citet{2018ApJ...858...31S} focuses on young stars with ages less than 1~Gyr (the median age is 50~Myr, with most being younger than 250~Myr), our selection targets massive stars with $M_\mathrm{ini}\geq 8M_\odot$, whose ages are at most $Age\lesssim 50~\mathrm{Myr}$ (see \secref{sec:prop}).
Secondly, there is a difference in selection strategy. \citet{2018ApJ...858...31S} utilized a wide color range of 46,148 samples to obtain a comprehensive representation of young stars, without accounting for false detections. In contrast, our approach prioritizes reducing the false detection rate rather than completeness (see \secref{sec:CMD}, \appref{sec:foreground}).
Lastly, the differences in the bandwidth of the Gaussian kernel also play a role. \citet{2018ApJ...858...31S} employed a bandwidth of 10~pc, which is feasible due to their large sample size of 46,148. In our case, using a 10~pc bandwidth with our sample of 7,426 would result in insufficient overlap of individual Gaussians, necessitating the use of a broader kernel function for a fair comparison against the 46,148 near-infrared samples.
Given these differences, it is likely that the discrepancies between the KDE map from \citet{2018ApJ...858...31S} and our \figref{KDE} arise primarily from variations in the stellar populations and resolution of KDE.

\subsection{Comparison with Comprehensive Spectroscopic Observation Catalog}
\label{sec:bonanos}

In this section, we demonstrate that the majority of the particularly massive stars from the massive star catalog of the SMC by \citet{2010AJ....140..416B} are selected using our massive star selection method outlined in \secref{sec:CMD}. Furthermore, we discuss the completeness of our catalog based on the number of particularly massive stars that were not selected.
\citet{2010AJ....140..416B} compiles massive stars identified in the SMC through past spectroscopic observations, reporting a total of 5,324 massive stars in the catalog (hereafter referred to as the B10 catalog).
The B10 catalog is the most comprehensive spectral catalog of massive stars in the SMC, widely cited (e.g., \citealp{2021A&A...646A.106S, 2024AJ....168..255H}).
\citet{2021A&A...646A.106S} created a Hertzsprung-Russell diagram for the B10 catalog using \textit{Gaia} DR2, concluding that completeness is $\approx$40\% for faint O-type stars at $G\approx15.5~\mathrm{mag}$, over 70\% for the brightest O-type stars at $G\lesssim14~\mathrm{mag}$, and nearly complete for the brightest stars, with any biases being negligible.
The uncertainty in the spectral type classifications of each star in the B10 catalog varies depending on the source study.
For example, the uncertainty in the spectral types from \citet{2004MNRAS.353..601E}, which constitute the majority of the B10 catalog, arises from factors such as limited sensitivity and the absence of emission lines used in spectral classification, with their classification being performed visually.
Due to its nature as a compilation of previous observations, the B10 catalog suffers from varying observational sensitivities across different regions, rendering it unsuitable for statistical analysis. Additionally, we point out that the B10 catalog will contain many stars with masses of $\mathrm{<}8M_{\odot}$.
In fact, \citet{2010AJ....140..416B} notes in footnote 15 of their paper that the objects identified in the observations by \citet{2004MNRAS.353..601E} (hereafter referred to as E04 stars) might include several intermediate- to low-mass stars. The E04 stars account for over three-quarters of the B10 catalog, totaling 4,073 objects\footnote{Stars referenced in the B10 catalog from observations more recent than \citet{2004MNRAS.353..601E} are not included in this count, while \citet{2004MNRAS.353..601E} lists 4,161 objects.}. To ensure a fair comparison, it is essential to first obtain reliable massive stars from the B10 catalog.

\begin{figure*}
 \centering
 \includegraphics[width=18cm,clip]{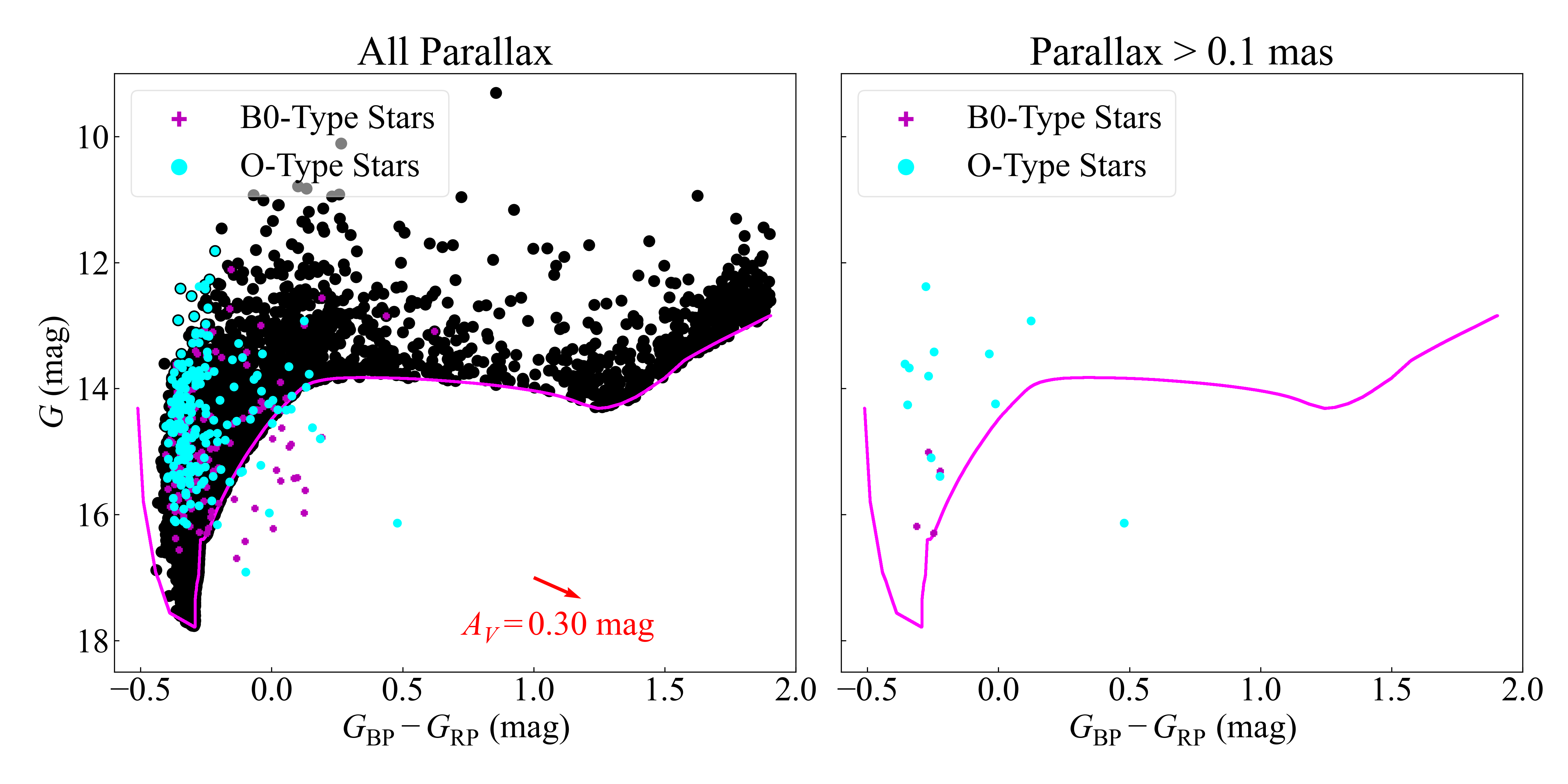} 
 \caption{
  (Left) The CMD of cross-matched B0-type and O-type stars in the B10 catalog, plotted as ($G_\mathrm{BP}-G_\mathrm{RP}$, $G$). The 130 B0-type stars are represented by purple crosses, and the 187 O-type stars are represented by light blue circles. The black circles represents the 7,426 massive star candidates identified by us. Below the magenta line delineating the boundary of massive stars, 18 B0-type stars and 13 O-type stars are plotted. The red arrow represents the reddening vector corresponding to $A_{V}=0.30~\mathrm{mag}$.
  (Right) Similar to the left panel, but for the CMD of stars with $\varpi > 0.1~\mathrm{mas}$, which were excluded from our data as foreground objects. Four B0-type stars and 13 O-type stars are plotted. Below the magenta line indicating the boundary of massive stars, there is one O-type star.
}
\label{B10_CMD}
\end{figure*}

\subsubsection{Selection of Reliable OB0-type Stars in the B10 Catalog}
\label{sec:B10OB0}

The upper left panel of \figref{B10_E04} shows all stars from the B10 catalog overplotted on a background of H$\alpha$ emission. The stars in the B10 catalog are distributed throughout the SMC and its outer region, exhibiting no distinct bar or wing structures, and their spatial distribution appears to differ from that of the background H$\alpha$ emission. However, most of the stars in the outer regions of the SMC from the B10 catalog are E04 stars, and examining the remaining 1,251 stars suggests that they are plotted in regions where H$\alpha$ is bright. The differing spatial distribution from the H\,\textsc{ii} regions expected to be created by massive stars is not surprising when considering that many intermediate- to low-mass stars are likely included among the E04 stars. Furthermore, the lack of separation between the wing and the southern part of the bar is consistent with the characteristics of stars older than $\sim$60~Myr \citep{2018MNRAS.478.5017R}.

The spectral types corresponding to masses of $\mathrm{\geq}8M_{\odot}$ are B2-type and earlier for main sequence stars (e.g., \citealp{2014ApJ...795...82S}). Therefore, we utilized the spectral type classification information provided by the B10 catalog to extract the most reliable candidates for stars exceeding $8M_{\odot}$, specifically 181 B0-type stars and 281 O-type stars.
In the extraction process, we removed stars that could potentially be classified as late-type due to uncertainties in spectral classification, specifically excluding B3-type stars and those that might be later types. The stars selected as B0-type include several possible B1-type and B2-type stars. To avoid duplication, stars that could be classified as both B0-type and O-type were counted as O-type\footnote{A single object with the classification \texttt{B0-O.5III-Ve XRB}, which is likely a typographical error due to the use of the letter ``O" instead of the numeral ``0," has been counted as a B0-type star.}.
The spatial distribution of stars classified as B0-type and O-type in the B10 catalog is shown in the upper right panel of \figref{B10_E04}. The spatial distribution of these stars aligns with the background H$\alpha$ emission, clearly exhibiting the bar and wing structures. This indicates that our massive star candidates also share a consistent spatial distribution with the B0-type and O-type stars. Therefore, we conclude that the extraction of B0-type and O-type stars has yielded the most reliable spectroscopically selected massive stars, which will be utilized in the following discussions. For discussions regarding other populations in the B10 catalog, please refer to \appref{sec:B10CMD}.

\begin{figure*}
 \centering
 \includegraphics[width=18cm,clip]{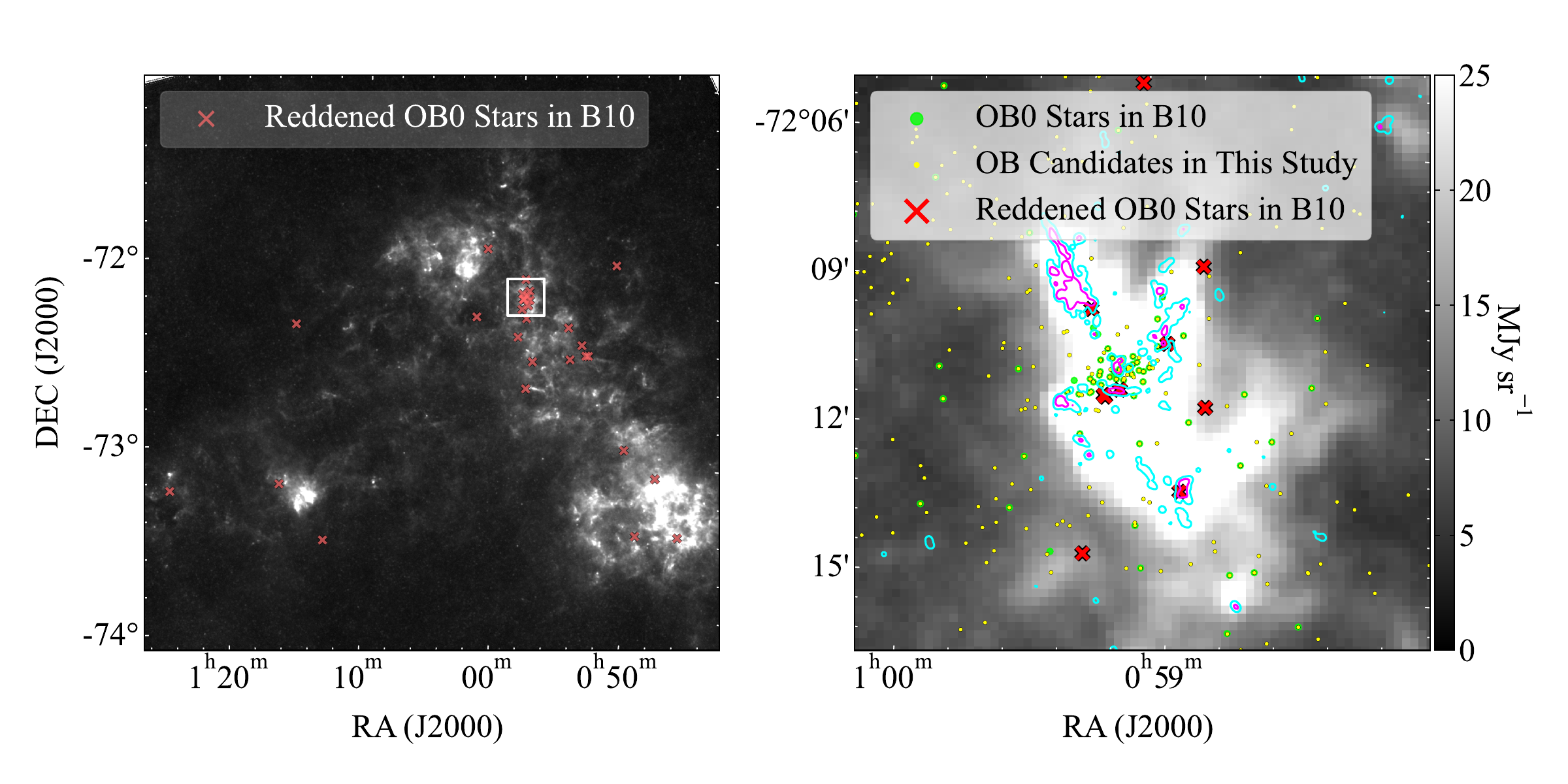}
 \caption{
  (Left) The spatial distribution of the reddened OB0 stars in the B10 catalog. The background shows the \textit{Herschel} $350~\mathrm{\text{\textmu}m}$ emission ranging from 0 to 25~$\mathrm{MJy~sr}^{-1}$. The red crosses represent the 31 stars located below the magenta line in \figref{B10_CMD}. The white square indicates the NGC~346 region, which is enlarged in the right panel.
  (Right) An enlarged view of the NGC~346 region indicated by the white square in the left panel. The contours represent the peak temperature map of CO~($J=2\text{--}1$) obtained with ALMA \citep{2021ApJ...922..171T, 2023ApJ...949...63O}. The contour levels are 3$\sigma$ for cyan and 10$\sigma$ for magenta. The red crosses again represent the 31 stars located below the magenta line in \figref{B10_CMD}, while the green circles represent the 287 OB stars from the B10 catalog that are positioned above the magenta line. The small yellow circles represent our massive star candidates.
  Note that, the massive star candidates plotted in this figure include many stars of later types than B0-type listed in the B10 catalog, as well as several newly detected as massive stars.
}
\label{B10_Red}
\end{figure*}

To verify that the B0-type and O-type stars from the B10 catalog are selected using our \textit{Gaia} CMD method, we need the $G~\mathrm{mag}$, $G_\mathrm{RP}~\mathrm{mag}$, and $G_\mathrm{BP}~\mathrm{mag}$ information for the stars in the B10 catalog. Therefore, we performed a cross-match between the B10 catalog and the all \textit{Gaia} DR3 dataset.
To ensure accuracy, we used a slightly small cross-match radius of $0.5''$, allowing us to find \textit{Gaia} DR3 counterparts for 2,465 out of the 5,324 stars in the B10 catalog\footnote{
With a cross-match radius of $1.0''$, we found 5,028 matches out of 5,324, and these results are also included in \tabref{catalog}.
}
\footnote{
\citet{2021A&A...646A.106S} used a radius of $3.0''$ and \textit{Gaia} DR2 to find 5,304 matches, but we note that using a $3.0''$ radius often includes multiple massive star candidates within the radius.
}.
The lower left panel of \figref{B10_E04} shows the spatial distribution of stars in the B10 catalog that successfully cross-matched. In comparison to the upper left panel of \figref{B10_E04}, there is a noticeable decrease in the number of E04 stars. Specifically, while the overall cross-match success rate is approximately 46\%, the success rate for E04 stars alone is approximately 32\%, which negatively impacts the cross-match success rate of the B10 catalog. When calculating the cross-match success rate for B10 catalog stars excluding E04 stars, the rate improves significantly to approximately 91\%.
To successfully cross-match E04 stars, it is necessary to increase the cross-match radius; however, this will lead to matches with incorrect objects. Therefore, we utilize a slightly smaller cross-match radius of $0.5''$ to ensure accuracy.

It is important to note that there are several instances where one \textit{Gaia} DR3 object is cross-matched with two B10 objects.
When using a cross-match radius of $0.5''$, there are two examples. Specifically, the B10 catalog object names \texttt{2dFS5100} and \texttt{NGC346-016} are cross-matched with the \textsc{source\_id} 4690503689148323328 in \textit{Gaia} DR3, as well as \texttt{SMC5\_037341} and \texttt{NGC330-110} are cross-matched with \textsc{source\_id} 4688993750489669120 in \textit{Gaia} DR3. This situation arises because the B10 catalog lists the same object multiple times. The source of the B10 catalog, \citet{2006A&A...456..623E}, reports that \texttt{NGC346-016} and \texttt{2dFS5100} are the same object, but this is not reflected in the B10 catalog. While the B10 catalog identifies \texttt{SMC5\_037341} and \texttt{NGC330-110} as the same object, \texttt{SMC5\_037341} is also listed again as a separate entity. Such duplications of objects in the B10 catalog also exist elsewhere\footnote{For example, \texttt{2dFS1037} and \texttt{NGC330-104} are also considered the same object in \citet{2006A&A...456..623E}. There appear to be instances in which the correspondence of the same objects in \citet{2006A&A...456..623E} and \citet{2004MNRAS.353..601E} is reflected in the B10 catalog, as well as instances where it is not.}, though the number of duplicates is at most around a dozen.
The 318 cross-matched B0-type and O-type stars in the B10 catalog include only \texttt{NGC346-016} as a B0-type star among the duplicated stars; duplication has no impact on the subsequent discussions.
However, researchers attempting to compare our catalog with the B10 catalog should be aware of this issue.
A detailed consideration of the multiple counts in the B10 catalog falls outside the scope of this paper, as no duplications, except for two cases, are included within the $0.5''$ cross-match radius we use, so we will not delve into it further.

The lower right panel of \figref{B10_E04} shows the spatial distribution of the 318 cross-matched B0-type and O-type stars from the B10 catalog. The cross-match success rates for B0-type and O-type stars are higher than the overall cross-match success rate for the B10 catalog, with 130 out of 181 B0-type stars and 188 out of 281 O-type stars identified in \textit{Gaia} DR3. This is due to the lower proportion of E04 stars among the B0-type and O-type samples. Although E04 stars comprise a significant portion of the B10 catalog, their representation among the extracted B0-type and O-type stars is relatively small, with 82 out of 181 and 132 out of 281, respectively. The cross-matched B0-type and O-type stars tend to avoid the outer regions of the galaxy, which can be attributed to the fact that most of the samples in the outer regions are dominated by E04 stars, which have low cross-match success rates. For the sake of convenience, we will refer to the 318 cross-matched B0-type and O-type stars in the B10 catalog simply as OB0 stars in the B10.

\subsubsection{Detection Rate of OB0-type Stars in the B10 Catalog}
\label{sec:RecovRate}
\paragraph{Reddening}
The left panel of \figref{B10_CMD} shows the ($G_\mathrm{BP}-G_\mathrm{RP}$, $G$) CMD of the 317 OB0 stars in the B10, excluding one O-type star for which $G_\mathrm{BP}-G_\mathrm{RP}$ is not available\footnote{The O-type star for which $G_\mathrm{BP}-G_\mathrm{RP}$ is not available is classified as spectral type O8V and is located in the NGC~346 region. Another brighter O-type star, classified as spectral type OC5V \citep{2000PASP..112.1243W}, is positioned at a distance of $1.94''$ from this star.}. Most of the plotted objects are located above the magenta line, although several stars are found below it. The number of stars plotted below the magenta line is 31, which accounts for approximately 10\% of the total OB0 stars in the B10. Hereafter, we refer to these 31 stars as reddened OB stars, as they are considered to be reddened within the SMC, thus falling outside our selection criteria.

The left panel of \figref{B10_Red} shows the spatial distribution of the 31 reddened OB0 stars in the B10. Most reddened OB stars are located within a bright filamentary dust structure at 350~\textmu m, showing a density excess in the northern region, specifically within the NGC~346 region, which is the largest star-forming region and contains significant amounts of dust. However, the presence of reddened OB0 stars in the bright dust regions can simply be explained by the overall abundance of OB0 stars in the NGC~346 region or by the alignment of their spatial distribution with that of the ISM. Therefore, we conduct a comparison with CO emission lines, which trace particularly dense ISM (a few times denser than the Milky Way, at $\sim$$10^{4}~\mathrm{cm^{-3}}$; \citealp{2017ApJ...844...98M, 2021ApJ...922..171T}). 

The right panel of \figref{B10_CMD} shows an enlarged view of the NGC~346 region along with contours of CO emission lines with ALMA.
In the NGC~346 region, despite the presence of a large number of our massive star candidates\footnote{In the KDE map shown in \figref{KDE}, using a Gaussian kernel bandwidth of 30~pc, NGC~346 appears as the second strongest peak, but with a smaller bandwidth, it emerges as the strongest peak. Thus, we can say that NGC~346 is also the region where our massive star candidates are distributed with the highest density.}, only a few overlap with the CO contours, indicating that they tend to avoid these CO contour regions and are instead distributed around them. Conversely, while the number of reddened OB stars is relatively small, several stars do overlap with the CO contours, making it plausible to conclude that the reddened OB stars are indeed distributed in regions of large reddening.
If the reddened OB stars are distributed in regions of large reddening, then stars that do not aline with the 350~\textmu m emission or CO contours can be interpreted as being locally surrounded by unresolved ISM.

An important point is that the area occupied by CO, which our massive star candidates avoid, is not large even in the NGC~346 region (see also \citealp{2021ApJ...922..171T}). CO is generally weak throughout the SMC, which is attributed to a lack of dust and low metallicity \citep{2001PASJ...53L..45M}. Our selection method covers most areas in the SMC where CO has not been detected. In regions where CO is detected, corresponding to several magnitudes of $A_{V}$, massive star progenitors should be detectable as young stellar objects in the infrared, complementing optical observations (e.g., \citealp{2023ApJ...949...63O,2024ApJ...971..108H}).

Note that, in addition to NGC~346, regions exhibiting the anti-correlation between 350~\textmu m and massive star candidates mentioned in \secref{sec:spacial} also contain CO~($J=2\text{--}1$) emissions. The northern region of NGC~330, the southwestern area, and the N83/N84 region located in the wing structure particularly exhibit a pronounced anti-correlation between massive star candidates and CO~($J=2\text{--}1$) emissions in archival data from ALMA (ID: 2018.1.01115.S and 2018.1.01319.S).

\paragraph{Parallax}
As mentioned in \secref{sec:data1}, there is a possibility that objects belonging to the SMC may be excluded from our selection due to an overestimation of their $\varpi$. The right panel of \figref{B10_CMD} shows the CMD of 17 OB0 stars in the B10 that have been excluded from our selection as foreground objects with $\varpi > 0.1~\mathrm{mas}$ (corresponding to a distance of $<$$10~\mathrm{kpc}$), which includes 4 B0-type stars and 13 O-type stars. This represents approximately 5.3\% of the total OB0 stars in the B10.
The parallax limitation does not lead to a significant loss in the sample and is smaller than the 10\% exclusion due to reddening.

Despite being brighter and expected to have more accurate parallax measurements, many O-type stars exhibit larger parallaxes compared to B0-type stars.
Even when considering the parallax error $\sigma_{\varpi}$, two of them (which are sufficiently bright; $G \sim 13~\mathrm{mag}$) still have $\varpi - \sigma_{\varpi} > 1~\mathrm{mas}$, suggesting the presence of non-negligible systematic errors in the parallax measurements. 
These stars may be influenced by interactions with other objects, such as in binary systems, leading to motion components distinct from annual motion and proper motion. Alternatively, if there are mistakes in the classification from the B10 catalog, these could potentially include late O-type stars in the outer regions of the Milky Way, located at distances greater than $\sim$10~kpc; however, there is no active evidence to support this.

\paragraph{Total Detection Rate}
Based on the above, the detection rate of OB0 stars in the B10 catalog within our sample is approximately 85\%, considering the 10\% excluded due to reddening and the 5\% excluded due to parallax limitations.
Therefore, most of the massive stars in the B10 catalog can be selected using our method.
It is unlikely that there are significant numbers of massive stars in the SMC not included among our candidates—whether due to reddening, parallax, or other factors—that would be comparable to the size of our sample.
This is supported by the fact that the majority of O-type stars in the central region of NGC~346 (central 5$\times$5 pixels in the right panel of \figref{B10_Red}), as listed by \citet{2022A&A...666A.189R}, are included in our catalog (see \secref{sec:3.2.5}).
By using the uniformly observed \textit{Gaia} DR3 data across the entire SMC, we conclude that we have obtained a statistically significant and larger sample of massive stars compared to the B10 catalog.

\begin{figure*}
 \centering
 \includegraphics[width=18cm,clip]{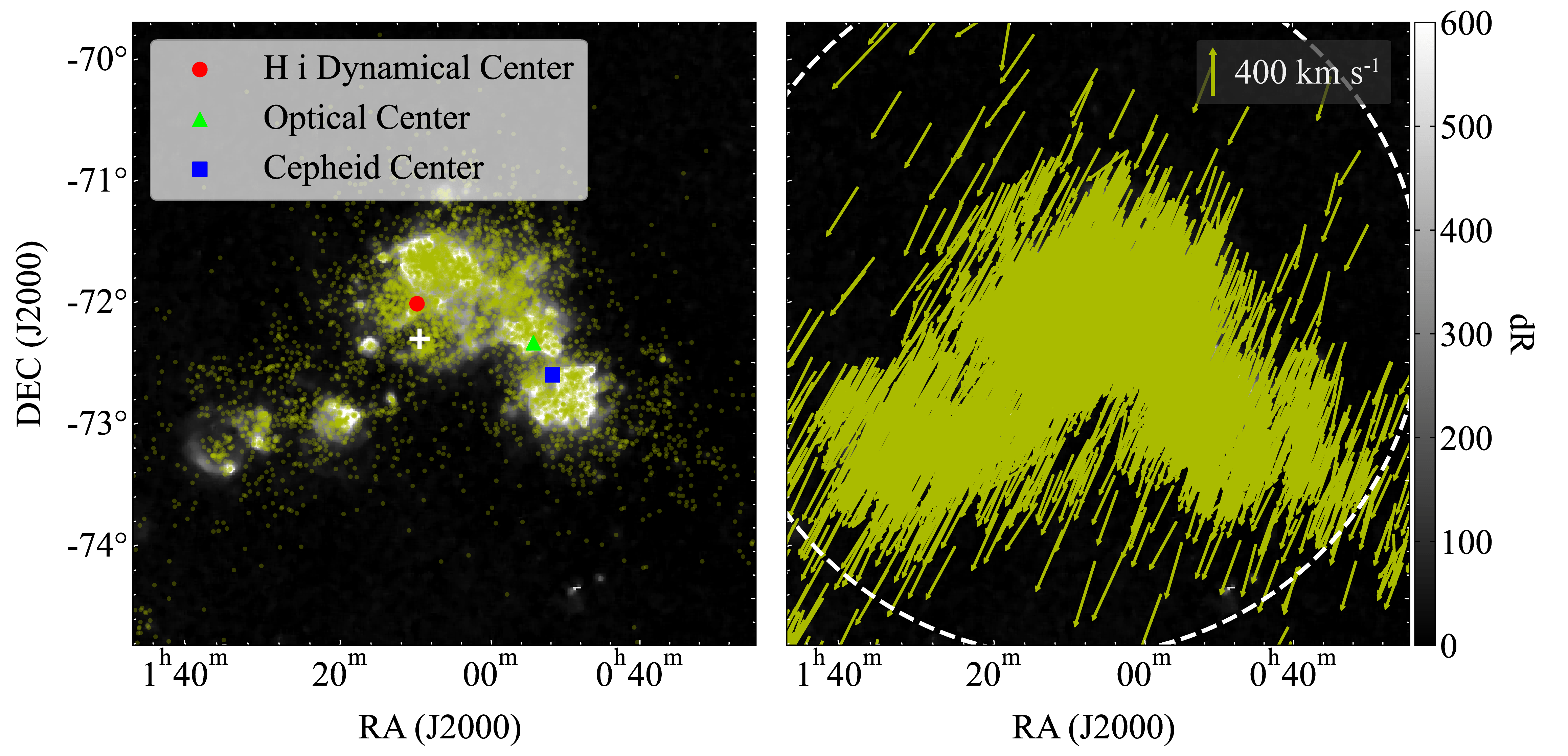}
 \caption{
  (Left) The spatial distribution of 7,261 massive star candidates with PM in both RA and DEC directions contained within 2$\sigma$ (the clipped samples). The white cross represents the mean position of all the 7,426 massive star candidates. The red circle represents the dynamical center of the H\,\textsc{i} gas \citep{2004ApJ...604..176S}, the green triangle represents the optical center \citep{1972VA.....14..163D}, and the blue square represents the center of the classical Cepheids distribution \citep{2017MNRAS.472..808R}.
  (Right) The PM vectors of the clipped samples. The legend indicates the magnitude of the PM vector corresponding to $400~\mathrm{km~s^{-1}}$, assuming a distance of 65~kpc. The white dashed circle represents a radius of $3^{\circ}$ centered on the dynamical center of the H\,\textsc{i} gas, which is used for calculating the systemic PM of the massive star candidates.
}
\label{SMC_PM}
\end{figure*}

\section{Motion of Massive Star Candidates}
\label{sec:PM}

In this section, we present the \textit{Gaia} PMs and RVs of the massive star candidates selected in \secref{sec:selection}. Our primary focus is on the internal motions of these massive star candidates within the SMC, which are expected to trace the dynamics of the ISM. To calculate the internal PMs, it is necessary to subtract the systemic PM of the SMC from the \textit{Gaia} PMs. However, since the spatial distribution of stars in the SMC varies by population, different populations may possess different systemic PMs. Therefore, we first calculate the systemic PM of the selected massive star candidates. The detailed procedure for calculating the internal PMs is outlined in \secref{sec:pm}, and the results of these calculations are presented in \secref{sec:pm_result}.
The properties of the RVs available for some of the massive star candidates, along with a comparison to the H\,\textsc{i} line-of-sight velocities, are discussed in \secref{sec:RV_OB}.

\subsection{Calculation of the Internal PM}
\label{sec:pm}

In this section, we calculate the internal PMs of the selected massive star candidates. Below, the PM in the RA direction, corrected for coordinate distortion by multiplying with $\mathrm{cos}\,\delta\,$, is denoted as $\mu_\alpha$, while the PM in the DEC direction is denoted as $\mu_\delta$. To discuss the global PM of the massive stars, it is essential to exclude dynamically ejected stars from star clusters. To remove runaway stars, we selected 7,261 massive star candidates that fall within $\pm2\sigma$ of both $\mu_{\alpha}$ and $\mu_{\delta}$.
Since stars moving at high velocities remain at the 1$\sigma$ limit, while stars such as those in the bridge region are excluded at the 3$\sigma$ limit, we adopt a 2$\sigma$ limit.
By performing sigma-clipping in the RA and DEC directions, we can remove the objects that deviate from the mean direction of motion, including few remaining foreground objects.
In the following, we refer to the PM 2$\sigma$-clipped massive star candidates as the clipped samples.
The left panel of \figref{SMC_PM} shows the spatial distribution of the clipped samples. The right panel of \figref{SMC_PM} shows the PM vectors of the clipped samples.

The mean PM of the clipped samples is $(\mu_\alpha,\,\mu_\delta)=(0.801,\,-1.251)~\mathrm{mas~yr^{-1}}$\,\footnote{This is simply the mean calculated for $\mu_\alpha$ and $\mu_\delta$ of the massive star candidates, without considering coordinate distortion.}, which is consistent with the systemic PM of the SMC obtained from the bar by \citet{2008AJ....135.1024P}, $(\mu_\alpha,\,\mu_\delta)=(0.754\pm0.061,\,-1.252\pm0.058)~\mathrm{mas~yr^{-1}}$, although $\mu_{\alpha}$ appears to be slightly larger.
We note that the magnitude of the mean PM of our clipped samples is $458.1~\mathrm{km~s^{-1}}$, and the mean errors of the PM are $(\sigma_{\mu_\alpha},\,\sigma_{\mu_\delta})=(14.4,\,13.3)~\mathrm{km~s^{-1}}$, both based on the assumption of a distance of 65~kpc.

PMs of the clipped samples are dominated by the systemic PM of the SMC. To obtain the internal PMs, we need to calculate the systemic PM of the clipped samples and subtract it from the PMs of our massive star candidates. Since the orientations in RA and DEC vary with the positions of the objects, the calculations of the systemic PM and internal PM must be conducted in isotropic coordinates. Taking into account the effects of coordinate distortion, we perform a transformation to an orthographic projection, similar to the method used by \citet{2018A&A...616A..12G}. Let the coordinates of the orthographic projection be $(x,\,y)$; the transformation is carried out as follows:
\begin{eqnarray}\label{3}
x &=& \mathrm{cos}\,\delta\,\mathrm{sin}(\alpha-\alpha_\mathrm{H_{I}}), \nonumber \\
y &=& \mathrm{sin}\,\delta\,\mathrm{cos}\,\delta_\mathrm{H_{I}}\,-\mathrm{cos}\,\delta\,\mathrm{sin}\,\delta_\mathrm{H_{I}}\,\mathrm{cos}(\alpha-\alpha_\mathrm{H_{I}}),
\end{eqnarray}

\begin{eqnarray}\label{4}
\mu_{x} &=& a\mu_{\alpha} - b\mu_{\delta}, \nonumber \\
\mu_{y} &=& c\mu_{\alpha} + d\mu_{\delta}.
\end{eqnarray}
Where,
\begin{eqnarray}\label{5}
a &=& \mathrm{cos}(\alpha- \alpha_\mathrm{H_{I}}), \nonumber \\
b &=& \mathrm{sin}\,\delta\,\mathrm{sin}(\alpha-\alpha_\mathrm{H_{I}}), \nonumber \\
c &=& \mathrm{sin}\,\delta_\mathrm{H_{I}}\,\mathrm{sin}(\alpha-\alpha_\mathrm{H_{I}}), \nonumber \\
d &=& \mathrm{cos}\,\delta\,\mathrm{cos}\,\delta_\mathrm{H_{I}} + \mathrm{sin}\,\delta\,\mathrm{sin}\,\delta_\mathrm{H_{I}}\,\mathrm{cos}(\alpha-\alpha_\mathrm{H_{I}}),
\end{eqnarray}
and $(\alpha_\mathrm{H_{I}},\,\delta_\mathrm{H_{I}})=(16.26^{\circ},\,-72.42^{\circ})$ is the dynamical center of the H\,\textsc{i} gas \citep{2004ApJ...604..176S}, corresponding to the center of the coordinate system used by \citet{2018A&A...616A..12G}.

Then, we calculate the systemic PM $(\mu_{x,\mathrm{sys}},\,\mu_{y,\mathrm{sys}})$ under the orthographic projection. 
Since many of the massive star candidates are located in the bridge, which is away from the SMC, we select the clipped samples situated near the center of the galaxy and adopt their mean PM as the systemic PM.
By calculating the mean PM of the clipped samples within 3$^{\circ}$ of $(\alpha_\mathrm{H_{I}},\,\delta_\mathrm{H_{I}})$, we obtain $(\mu_{x,\mathrm{sys}},\,\mu_{y,\mathrm{sys}}) = (0.785,\,-1.245)~\mathrm{mas~yr^{-1}}$. Note that at $(\alpha_\mathrm{H_{I}},\,\delta_\mathrm{H_{I}})$, $(\mu_{x},\,\mu_{y}) = (\mu_\alpha,\,\mu_\delta)$, which means that $(\mu_{x,\mathrm{sys}},\,\mu_{y,\mathrm{sys}})$ is equal to $(\mu_{\alpha,\mathrm{sys}},\,\mu_{\delta,\mathrm{sys}})$.

Using the calculated systemic PMs $(\mu_{x,\mathrm{sys}},\,\mu_{y,\mathrm{sys}})$, the internal PMs $(\mu_{x,\mathrm{in}},\,\mu_{y,\mathrm{in}})$ under the orthographic projection can be calculated as follows:
\begin{eqnarray}\label{6}
\mu_{x,\mathrm{in}} &=& \mu_{x} - \mu_{x,\mathrm{sys}}, \nonumber \\
\mu_{y,\mathrm{in}} &=& \mu_{y} - \mu_{y,\mathrm{sys}}.
\end{eqnarray}
The internal PMs $(\mu_{\alpha,\mathrm{in}},\,\mu_{\delta,\mathrm{in}})$ in the equatorial coordinate system can be obtained from the following coordinate transformation:
\begin{eqnarray}\label{7}
\mu_{\alpha,\mathrm{in}} &=& \frac{d\mu_{x,\mathrm{in}} + b\mu_{y,\mathrm{in}}}{ad+bc}, \nonumber \\
\mu_{\delta,\mathrm{in}} &=& \frac{-(c\mu_{x,\mathrm{in}} - a\mu_{y,\mathrm{in}})}{ad+bc}.
\end{eqnarray}
Through the calculations above, the internal PMs of the massive star candidates have been obtained.

The white cross shown in the left panel of \figref{SMC_PM} represents the mean position of our 7,426 massive star candidates ($\mathrm{RA}=01\mathrm{h}\,\,03\mathrm{m}\,\,14.26\mathrm{s}$, $\mathrm{DEC}=-70^{\circ}\,\,50'\,\,55.6''$). This mean position is located at the junction of the bar and wing, in close proximity to the dynamical center of the H\,\textsc{i} gas, and is notably distant from both the optical center, and the Cepheid center. This is consistent with the idea that the young massive star candidates trace the ISM more effectively than older stars, thereby supporting the validity of our selection. Furthermore, this is in agreement with the findings of \citet{2018MNRAS.478.5017R}, which suggest that recent star formation is occurring north of the bar.
Therefore, we used the dynamical center of the H\,\textsc{i} gas for the calculation of the systemic PM. As shown by the dashed circle in the right panel of \figref{SMC_PM}, the area within a radius of 3$^{\circ}$ centered on the dynamical center of the H\,\textsc{i} gas encompasses almost all of the massive star candidates in the SMC, excluding those in the bridge.

\begin{figure*}
 \centering
 \includegraphics[width=18cm,clip]{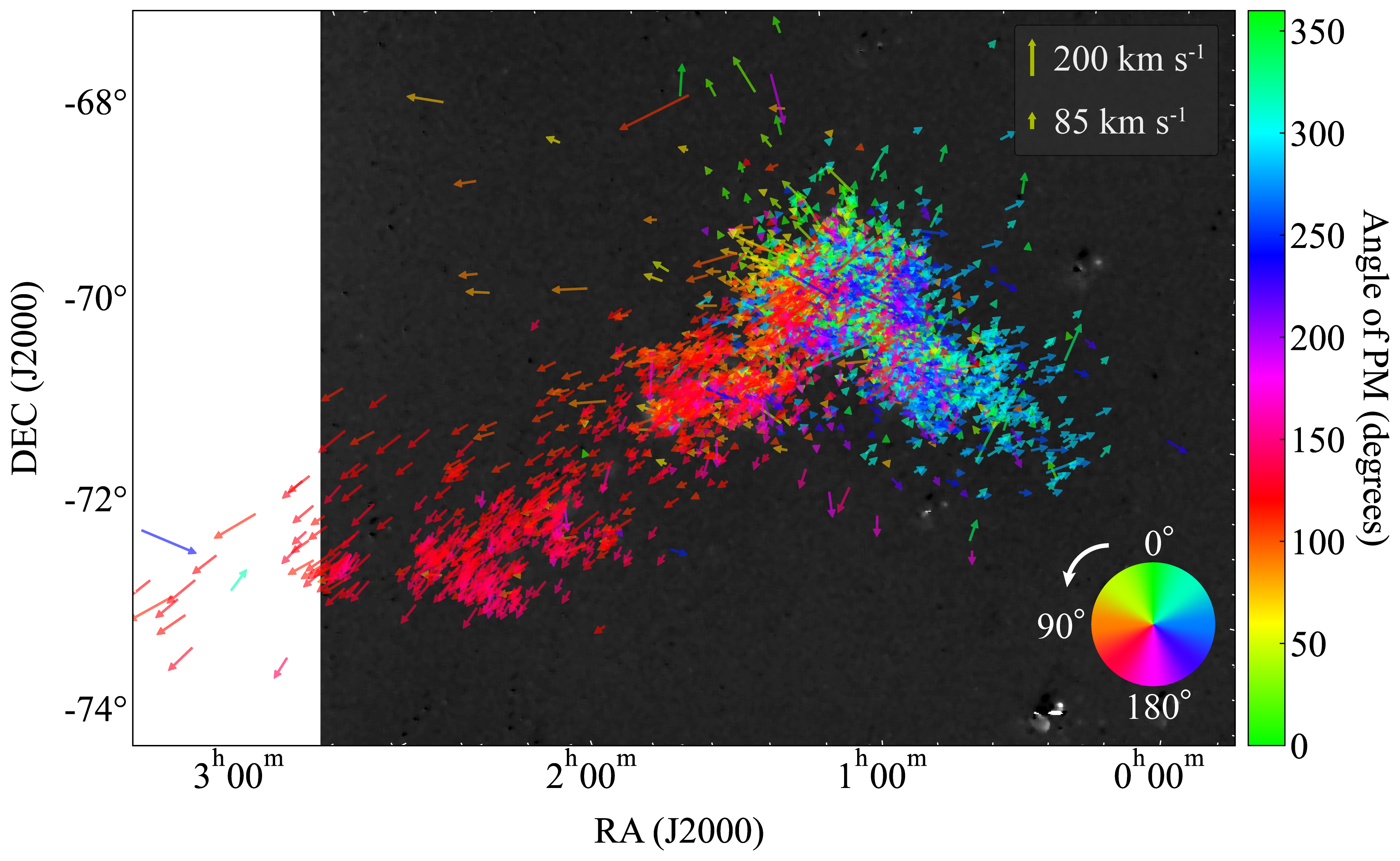}
 \caption{
  The internal PMs of the clipped samples determined by subtracting the mean PM of the clipped samples within 3$^{\circ}$ of the H\,\textsc{i} dynamical center. To account for the effects of coordinate distortion during this subtraction, a coordinate transformation used by \citet{2018A&A...616A..12G} is applied, and the proper motion is subtracted using the transformed coordinates. The color of the PM vectors represents the direction of the PM vector, and the angle is defined counterclockwise from north (0°) in the ($x, y$) coordinates based on the orthographic projection defined in \secref{sec:pm}. The legends indicate the magnitudes of PM vectors corresponding to $200~\mathrm{km~s^{-1}}$ and $85~\mathrm{km~s^{-1}}$, assuming a distance of 65~kpc, with the latter representing the minimum escape velocity derived under the assumption of H\,\textsc{i} rotation \citep{2004ApJ...604..176S,2018NatAs...2..901M}.
}
\label{SMC_inPM}
\end{figure*}

\begin{figure*}
 \centering
 \includegraphics[width=18cm,clip]{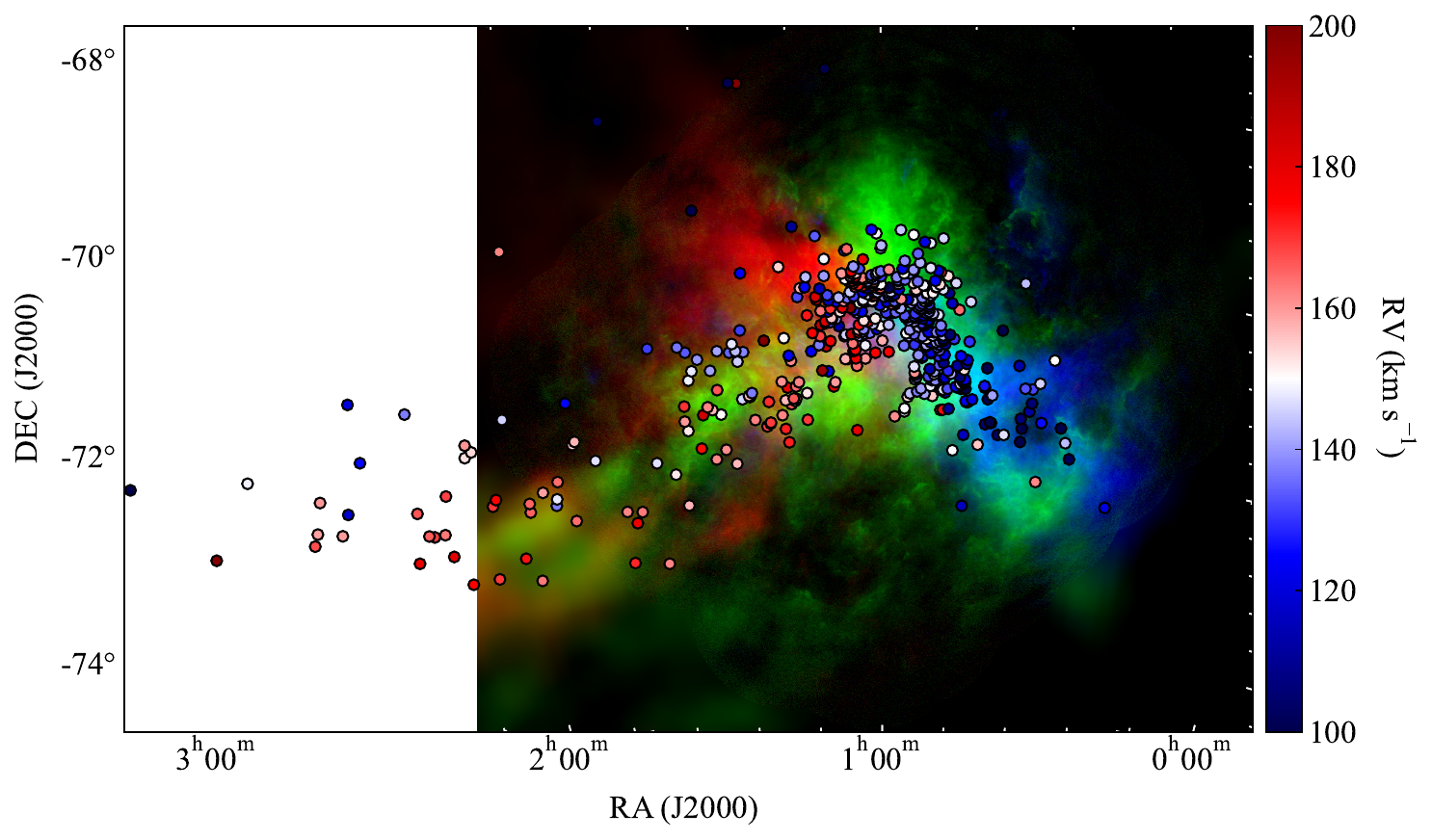}
 \caption{
  Comparison of the RVs of the clipped samples with the three velocity components of H\,\textsc{i}. The overplotted circles represent the 491 objects within the clipped samples for which RVs are available. The color of the circles indicates the magnitude of the RVs.
  The background is a color multivelocity montage of the H\,\textsc{i} emission with ASKAP \& Parkes \citep{2018NatAs...2..901M}, with red representing the integrated intensity for the 179--187$~\mathrm{km~s^{-1}}$ range, green for the 144--152$~\mathrm{km~s^{-1}}$ range, and the blue for the 105--113$~\mathrm{km~s^{-1}}$ range, all on a common scale of 10--800$~\mathrm{K~km~s^{-1}}$ (log scale).
  Note that these velocity components are consistent with those shown in Figure~2 of \citet{2018NatAs...2..901M}.
  The line-of-sight velocities of both H\,\textsc{i} and the clipped samples are referenced to the LSRK.
}
\label{SMC_RV}
\end{figure*}

It is important to be cautious of the fact that the calculation implicitly assumes all the massive star candidates are at the same distance.
If the distance to each individual massive star candidate is taken into account, equation \eqref{6} needs to be rewritten as follows:
\begin{eqnarray}\label{8}
\mu_{x,\mathrm{in}} &=& \mu_{x} - \frac{D_\mathrm{SMC}}{D_\mathrm{star}}\mu_{x,\mathrm{sys}}, \nonumber \\
\mu_{y,\mathrm{in}} &=& \mu_{y} - \frac{D_\mathrm{SMC}}{D_\mathrm{star}}\mu_{y,\mathrm{sys}},
\end{eqnarray}
where $D_\mathrm{SMC}$ is the distance to the SMC and $D_\mathrm{star}$ is the distance to each star.
Thus, the direction of the internal PMs can vary depending on the ratio of $D_\mathrm{SMC}$ to $D_\mathrm{star}$. Specifically, considering distance, the direction of the internal PM will shift toward the north-northwest for closer stars and toward the south-southeast for more distant stars, since the systemic PM of the SMC is directed south-southeast with a magnitude of $\sim$400~$\mathrm{km~s^{-1}}$ (see \figref{SMC_PM}).
This effect cannot be ignored when dealing with massive star candidates that have a line-of-sight depth of $\sim$20--30~kpc, as represented by classical Cepheids \citep{2017MNRAS.472..808R}, and accurate estimates of $D_\mathrm{SMC}$ and $D_\mathrm{star}$ are required to compute accurate internal PMs.

Conversely, when converting PM from units of $\mathrm{mas~yr^{-1}}$ to tangential velocity in units of $\mathrm{km~s^{-1}}$, the value of $D_\mathrm{SMC}$ affects only the magnitude of the PM vector, as the tangential velocity is obtained by multiplying $\mu~\mathrm{(mas~yr^{-1})}$ by $4.7438 \times D_\mathrm{SMC}~\mathrm{(kpc)}$.
Here, we again apply the assumption that all stars are located at a distance $D_\mathrm{SMC}$.
For conversions to velocity, which do not affect the direction of the PM vector, our rough estimate of $D_\mathrm{SMC}=D_\mathrm{star}=65~\mathrm{kpc}$ has little impact on the subsequent discussion.
Even when considering extreme values of $D_\mathrm{star}$ such as 50~kpc or 80~kpc, the difference from $D_\mathrm{SMC}$ is less than 25\%.


\subsection{Result of the Internal PMs Calculation}
\label{sec:pm_result}

\figref{SMC_inPM} shows the internal PMs of the clipped samples calculated in \secref{sec:pm}. 
The massive star candidates in the outer regions exhibit motions moving away from the main body of the SMC, consistent with previous studies (e.g., \citealt{2018ApJ...864...55Z,2021MNRAS.502.2859N}). 
The internal PMs trend of the massive star candidates is similar to that of the 143 young stars shown by \citet{2019ApJ...887..267M}, and similarly, it does not exhibit a clear galactic rotation.
The massive star candidates moving toward the east-northeast, away from the SMC, are consistent with the features observed in H\,\textsc{i} \citep{2003MNRAS.339..105M,2007MNRAS.381L..11M}, as well as those moving north may be associated with gas related to the Magellanic Stream.
As indicated by \citet{2018ApJ...867L...8O}, stars in the wing are moving toward the LMC, and stars in the bridge exhibit similar behavior. These stars are moving away from the SMC at velocities exceeding the minimum escape velocity of $85~\mathrm{km~s^{-1}}$ derived under the assumption of H\,\textsc{i} rotation \citep{2004ApJ...604..176S,2018NatAs...2..901M}. This suggests we may be witnessing the process by which the SMC loses mass due to tidal interactions and/or feedback mechanisms.

The motion of stars in the wing and bridge, when traced back in time,
returns to the northern region of the SMC. This supports the validity of establishing the center of the massive star candidates at the H\,\textsc{i} center.
The motion of stars in the bridge, moving at $\sim$100~$\mathrm{km~s^{-1}}$, suggests they were in the northern part of the SMC 50~Myr ago, exceeding the lifetime of massive stars and supporting in-situ star formation in the bridge (e.g., \citealp{2015MNRAS.450..552P}).
We note that when calculating the internal PMs using different systemic PMs—such as near the optical center or the midpoint of the H\,\textsc{i} center \citep{2004ApJ...604..176S} and the Cepheid center \citep{2017MNRAS.472..808R} as utilized by \citet{2018ApJ...867L...8O}—the stars in the wing, bridge, and western part of the SMC appear to have originated from a region devoid of any radiation to the south of the SMC.

\subsection{Properties of the RVs}
\label{sec:RV_OB}
\textit{Gaia} DR3 provides RV measurements for sufficiently bright stars (with $G \lesssim 14~\mathrm{mag}$), allowing some of the massive star candidates in the SMC to have their RVs available as line-of-sight velocities \citep{2023A&A...674A...5K}.
In this section, we show the line-of-sight motion of the massive star candidates using RVs available for some of them, and compare it with the velocity structure of H\,\textsc{i}.
Among our massive star candidates, RVs have been obtained for 588 stars, and for the clipped sample, 491 stars. These are composed of evolved stars with $G_\mathrm{BP}-G_\mathrm{RP} \gtrsim 0$.
The RV mean error of the clipped sample is $\sigma_\mathrm{RV}=1.85~\mathrm{km~s^{-1}}$.
Regarding the surface temperature \texttt{rv\_template\_teff} and magnitude \texttt{grvs\_mag} assumed during the RV calculations, \citet{2023A&A...674A...7B, 2023A&A...674A..32B} argue that for stars satisfying $8500\leq$ \texttt{rv\_template\_teff} $\leq14500$~K and $6\leq$ \texttt{grvs\_mag} $\leq12$~mag, a correction of \texttt{7.98 - 1.135 * grvs\_mag} should be subtracted from the RV. We applied the RV correction from \citet{2023A&A...674A...7B} to the four stars within our massive star candidates that required this adjustment.

\figref{SMC_RV} compares the three velocity components of H\,\textsc{i} (a blueshifted component at approximately 100$~\mathrm{km~s^{-1}}$, a systematic velocity component around 150$~\mathrm{km~s^{-1}}$, and a redshifted component at approximately 200$~\mathrm{km~s^{-1}}$) with the RVs of the clipped samples. For comparison in \figref{SMC_RV}, we convert the RVs of the clipped samples from the solar system barycentric reference frame to the LSRK\footnote{With the conversion to the LSRK, the RVs of the massive star candidates decrease by $\approx$11~$\mathrm{km~s^{-1}}$.}. 
From \figref{SMC_RV}, the RVs of the clipped samples exhibit a gradient in the northwest-southeast direction, with a higher number of stars having larger RVs in the wing located to the southeast.
The RV gradient in the northwest-southeast direction becomes more pronounced when looking at the RV distribution of slightly older stars that are also distributed in the southern SMC, as shown by \citet{2024ApJ...962..120M} for 1801 red supergiants with ages $\lesssim100~\mathrm{Myr}$ based on \textit{Gaia} DR3.
The RV velocity gradient in the northwest-southeast direction is consistent with the feature observed in independently measured RVs of E04 stars \citep{2008MNRAS.386..826E}\footnote{\citet{2008MNRAS.386..826E} obtained RVs for 2,045 out of 4,161 stars, most of which do not have RVs available in the \textit{Gaia} DR3.} and old or intermediate-age red giant stars \citep{2014MNRAS.442.1663D}.
These RVs of the clipped samples span a range similar to that of the H\,\textsc{i} velocities.
Notably, they align with the blueshifted regions of the bar. 
Similarly, in the wing, due to the presence of multiple velocity components of H\,\textsc{i}, it is possible that they align with the redshifted component as well as other components. 
In the northeastern part of the SMC, where a prominent H\,\textsc{i} redshifted component is noticeable, there are few massive star candidates, and the few tend to have small RVs.

The H\,\textsc{i} in the southwestern region of the SMC, colored in green and blue, appears a shape resembling a horn that extends towards the northwest, and massive star candidates in this region are also moving toward the northwest (see \figref{SMC_inPM}). Along the northwestern extension of this horn-like structure, the H\,\textsc{i} further extends in a filamentary manner \citep{2003ApJ...586..170P, 2005A&A...432...45B}, forming one side of the triangular structure known as the Magellanic Horn, as named by \citet{2012ApJ...750...36D}. 
The motion of the massive star candidates, whose RV coincide with the H\,\textsc{i} velocity, towards the northwest is consistent with the interpretation that this horn-like structure appears to have been stripped from the main body of the SMC.

Along the wing and the bridge in its extension, stars with small RVs are distributed in the northern part, while stars with larger RVs are found in the southern part. The characteristics of the RVs were also observed in the individual RV measurements of E04 stars by \citet{2008MNRAS.386..826E}, and are not attributable to systematic errors in \textit{Gaia}. The reason for this north-south RV separation is unclear, but it is possible that we are observing components at different depths, such as the bi-modal distance distribution found in the eastern SMC \citep{2017MNRAS.467.2980S, 2021MNRAS.504.2983T}.
The fact that the components with larger RVs in the southern part of the bridge have internal PMs directed relatively toward the south (\figref{SMC_inPM}) supports the hypothesis that the different RVs in the north and south in the bridge are due to the projection of distance.
This is because, for nearby stars, PM vectors are expected to point toward the south-southeast as a result of the residual systemic PM (see \secref{sec:pm}).
Alternatively, because the H\,\textsc{i} component with velocities less than 150~$\mathrm{km~s^{-1}}$ is thinly and broadly distributed at least up to the eastern region of the SMC observed with ASKAP \& Parkes, stars with small RVs may be associated with this component, which may not necessarily indicate a difference in distance.

\begin{figure}
 \centering
 \includegraphics[width=8.5cm,clip]{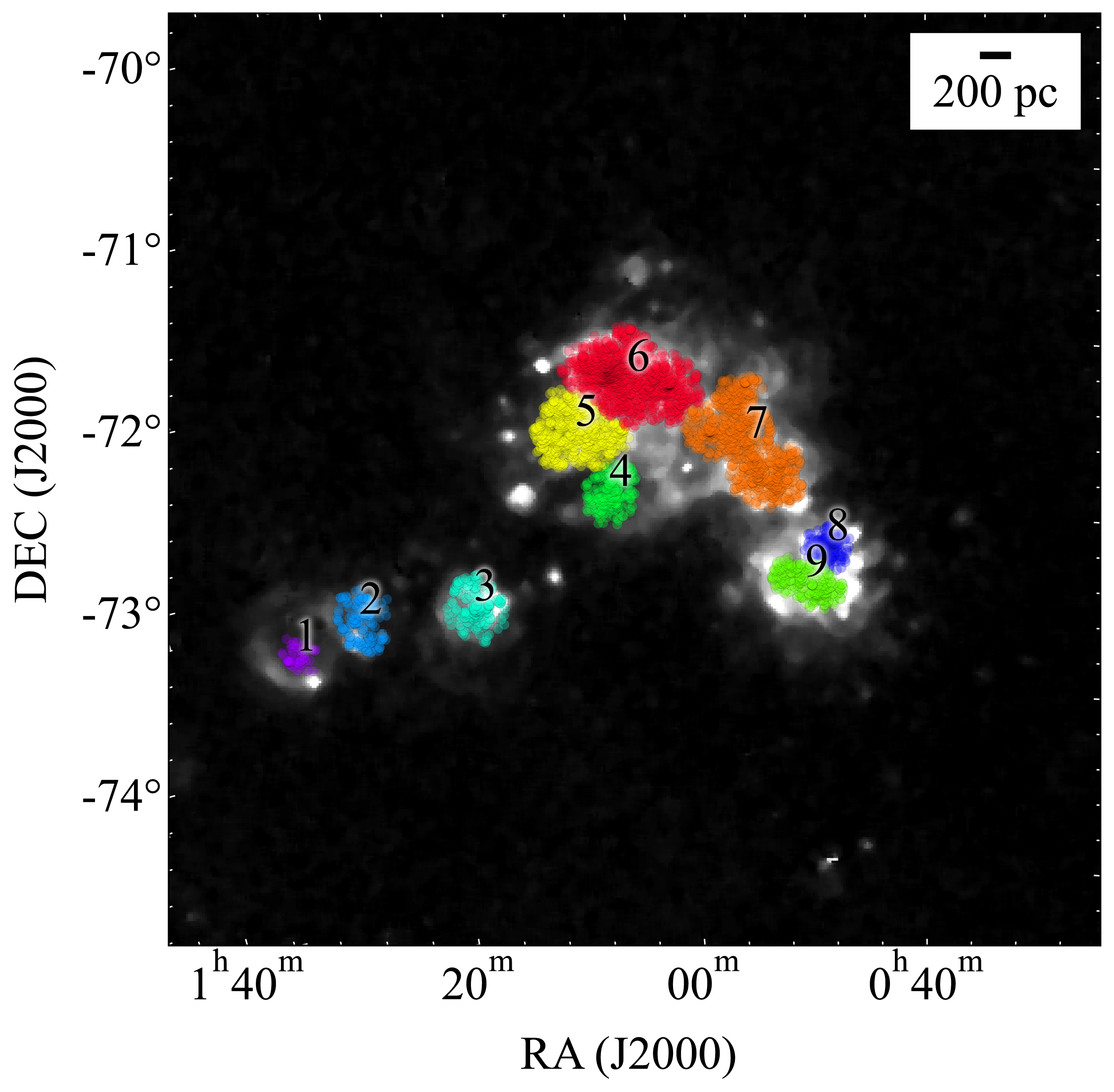}
 \caption{
  Nine superstructures identified using \texttt{DBSCAN}. Different structures are distinguished by color and numbers.
}
\label{SMC_st9}
\end{figure}

\begin{figure*}
 \centering
 \includegraphics[width=18cm,clip]{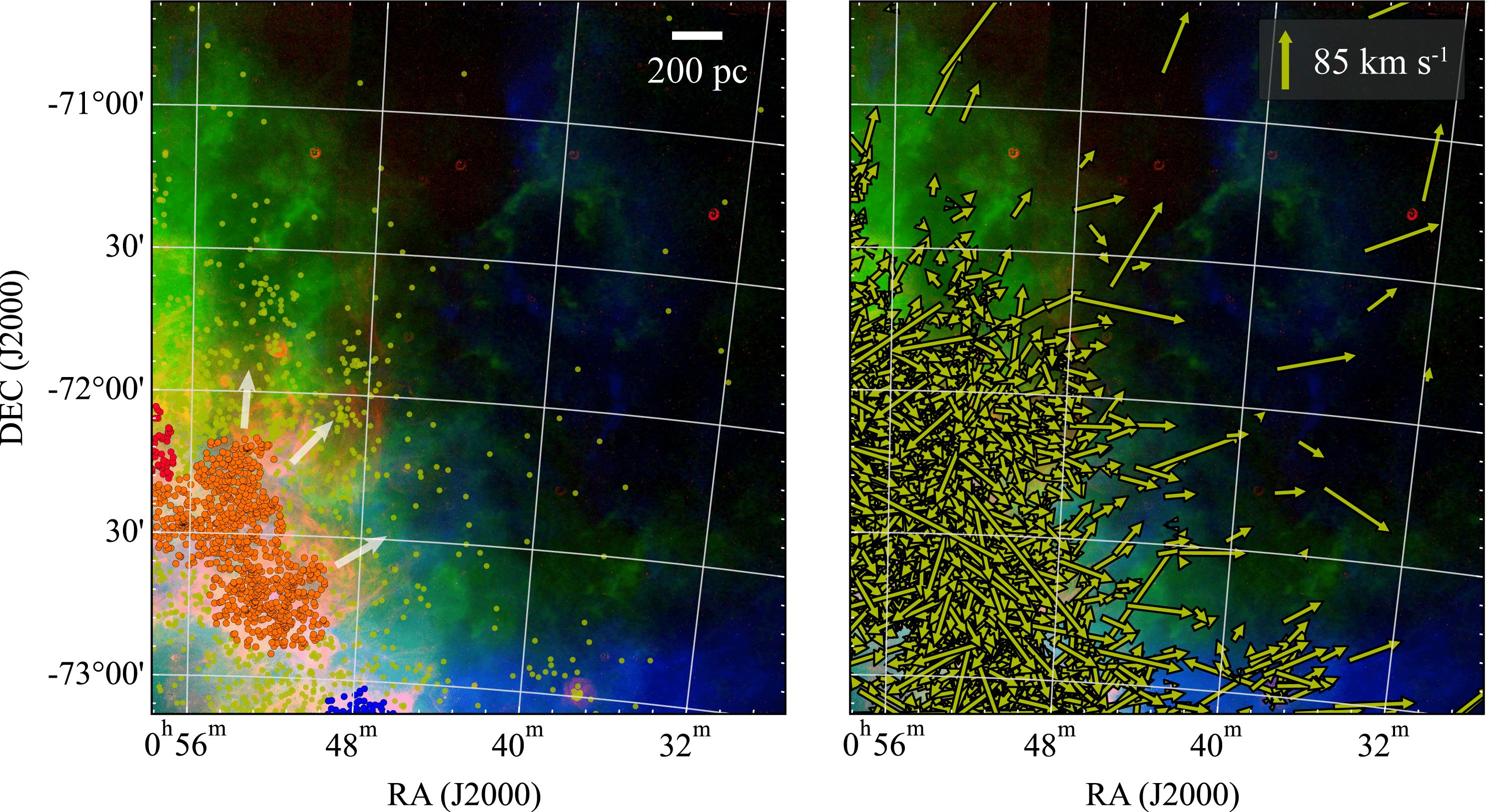}
 \caption{
  (Left) An enlarged view of the spatial distribution of the massive star candidates in the northwest of the SMC. The dark yellow circles represent massive star candidates. The superstructures are distinguished by circles of different colors. The white arrows highlight the lines of massive star candidates extending from the superstructure~7. The background is a color multiwavelength montage, showing the H$\alpha$ emission by MCELS (red; 5--$100\times10^{-17}~\mathrm{ergs~cm^{-2}~s^{-1}}$; \citealp{1999IAUS..190...28S}) and the H\,\textsc{i} emission with ASKAP \& Parkes (green and blue; integrated over the LSRK velocity range 144--152~$\mathrm{km~s^{-1}}$ and 105--113~$\mathrm{km~s^{-1}}$, respectively; 10--1200~$\mathrm{K~km~s}^{-1}$; \citealp{2018NatAs...2..901M}), all in log scale. The green and blue H\,\textsc{i} velocity components are the same as those shown in \figref{SMC_RV}.
  (Right) The internal PM vectors of the clipped samples in the region shown in the left panel. The background is the same as in the left panel. The legends indicate the magnitudes of PM vectors corresponding to $85~\mathrm{km~s^{-1}}$, representing the minimum escape velocity derived under the assumption of H\,\textsc{i} rotation.
}
\label{st7_nw}
\end{figure*}

\begin{figure*}
 \centering
 \includegraphics[width=12cm,clip]{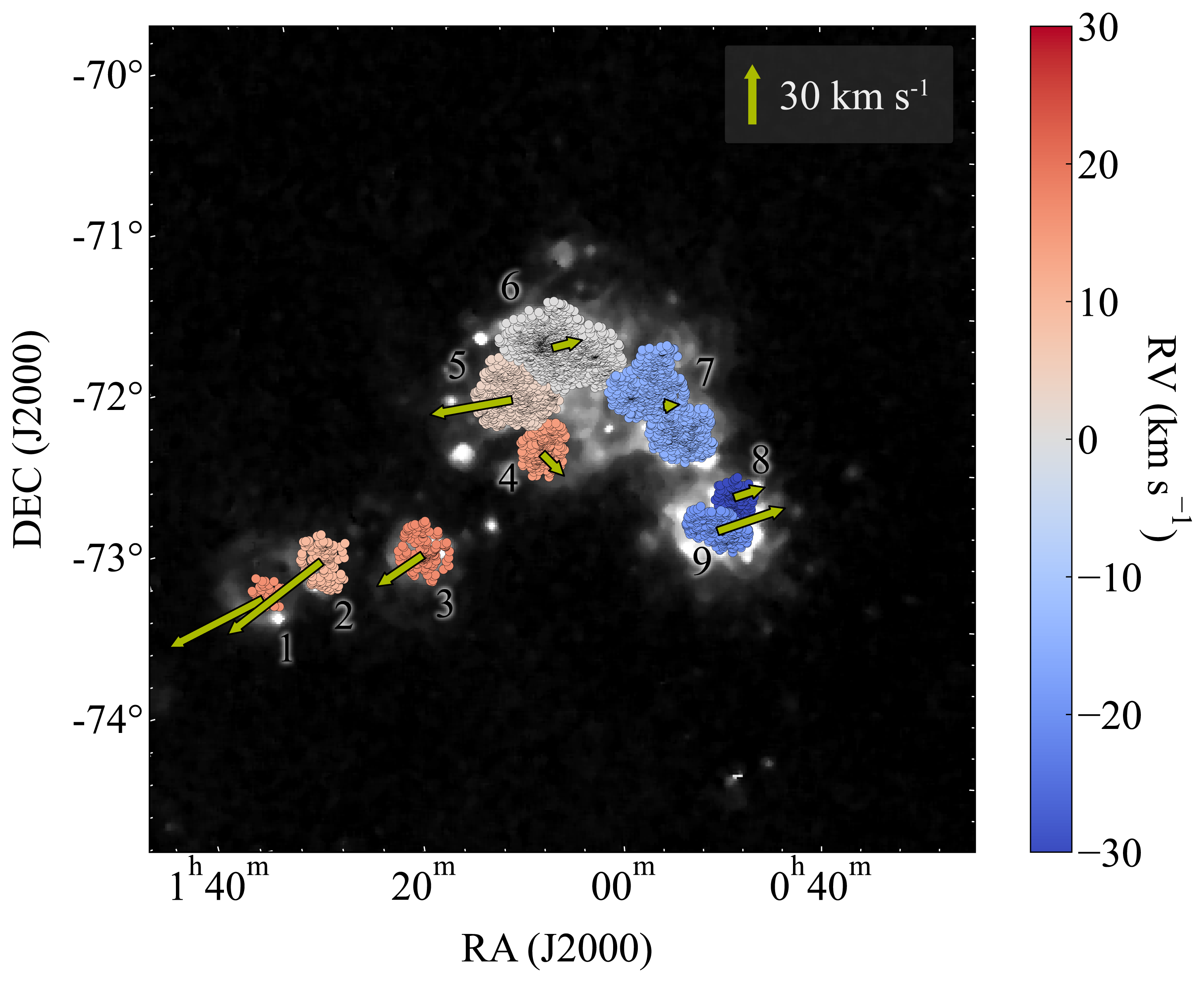}
 \caption{
  The median internal PMs and the mean RVs of the nine superstructures identified by \texttt{DBSCAN}. The vectors extending from each superstructure represent the median internal PMs, with the scale represented by the legend corresponding to a PM vector magnitude of $30~\mathrm{km~s^{-1}}$. The colors of each superstructure represent the mean RVs after subtracting the systemic RV of the SMC, which is $160~\mathrm{km~s^{-1}}$.
}
\label{SMC_st9_PM}
\end{figure*}

\section{The Superstructure Formed by Massive Star Candidates}
\label{sec:dbscan}
In this section, we perform a clustering analysis to separate several large-scale structures from the 7,426 massive star candidates. From physical properties such as median PM, we characterize the properties of the separated structures, which are composed of recently formed stars in the LMC-SMC interaction. Here, we refer to structures composed of massive stars on a scale of a few hundred parsecs as ``superstructures."

\subsection{Identification of the Superstructures}
\label{sec:identification_st}
As discussed in \secref{sec:spacial}, \figref{KDE} reveals several structures based on the surface density of the massive star candidates.
In this section, we identify superstructures to replicate the structures labeled North~1, North~2, and North~3 as shown in \figref{KDE}.
We use the density-based clustering algorithm \texttt{DBSCAN} \citep{1996kddm.conf..226E} for the identification of superstructures. The algorithm has two free parameters: the distance parameter $\epsilon$ (which corresponds roughly to the typical size of the structures) and the minimum points parameter $minPts$ (which corresponds roughly to the minimum number of members in a structure). Using these two input parameters, \texttt{DBSCAN} identifies structures that contain at least $minPts$ stars within a radius of $\epsilon$.

Although it is arbitrary, to identify North~1, North~2, and North~3 as distinct structures, we set $\epsilon = 100~\mathrm{pc}$ and $minPts = 80$ for the identification of superstructures in the bar, which corresponds to a density threshold of $\geq$$2.5\times10^{-3}~\mathrm{pc}^{-2}$.
For the identification of superstructures in the wing, we set $\epsilon = 100~\mathrm{pc}$ and $minPts = 40$, which is simply set as half of the density threshold for the bar. The use of a lower density threshold in the wing considers that the gravitational potential is weaker in this region, making it easier for massive stars to be scattered.

\figref{SMC_st9} shows the identified superstructures. A total of six superstructures are identified in the bar and three in the wing. Hereafter, we will refer to the identified superstructures using the labels 1--9, as shown in \figref{SMC_st9}.
The size of the superstructures is $\sim$100~pc, and when simply summing the masses of included massive star candidates ($\geq$$8M_{\odot}$), they contain at least $\sim$$10^{3\text{--}4}M_\odot$.
The superstructures~1, 2, and 3 are located in the wing, with the smaller numbered structures positioned further from the main body of the SMC and containing fewer massive star candidates.
The superstructures~1 and 2 associated with SMC-SGS1, the only known supergiant shell in the SMC. The superstructure~1 consists of the excess density within SMC-SGS-1, including NGC~602c and its surroundings, which hosts the SMC's only WO-type star. The superstructure~2 is situated in the western outer shell of SMC-SGS-1, comprising N88 and N89 and their surroundings. The superstructure~3 corresponds to N83 and N84, but the massive star candidates are more numerous slightly to the east of these regions.
The superstructure~4 is a superstructure of the bar that is closest to the wing, corresponding to North~3 in \figref{KDE}.
The superstructure~5, along with the superstructure~4, is situated at the junction of the bar and the wing, corresponding to North~2 in \figref{KDE}. As shown in \figref{3color}, the superstructures~4 and 5 exhibit an anti-correlation with the ISM.
The superstructure~6 is the densest structure in the northern bar, corresponding to North~1 in \figref{KDE}. According to \figref{3color}, the superstructure~6 shows a correlation with the ISM\footnote{Within the superstructure~6, there are variations in the intensity of the H\,\textsc{i} $\sim$$150\mathrm{~km~s^{-1}}$ component, with relatively high concentrations observed in the center and northern regions of NGC~371, as well as in the southern part of NGC~346. Additionally, relatively diffuse H\,\textsc{i} is seen around each massive star candidate, which is a general trend observed throughout the entire SMC.} and appears to be separated from the superstructure~5 by a boundary with fewer massive star candidates.
The superstructure~7 represents the main body of the bar and may further decompose depending on the parameters used.
\figref{st7_nw} shows an enlarged view of the spatial distribution of the massive star candidates and the internal PMs of the clipped samples around the superstructure~7.
The superstructure~7 has an angular shape, particularly noticeable as it extends northwest from the northern part of the structure along the H$\alpha$ that is anti-correlated with the  H\,\textsc{i} $\sim$$150\mathrm{~km~s^{-1}}$ component, roughly at ($\mathrm{RA}=00\mathrm{h}\,\,54\mathrm{m}$, $\mathrm{DEC}=-72^{\circ}\,\,00'$). As highlighted by the white arrows in the left panel of \figref{st7_nw}, to the north and northwest of this angular structure, there are lines of massive star candidates, which are indeed moving northwest. Furthermore, a line of massive star candidates moving northwest is also found extending from a protrusion located to the southwest of the superstructure~7, approximately at ($\mathrm{RA}=00\mathrm{h}\,\,50\mathrm{m}$, $\mathrm{DEC}=-72^{\circ}\,\,30'$).
The superstructures~8 and 9 coincide with the brightest regions of the ISM in the southwest; however, the superstructure~9 exhibits an anti-correlation with the filamentary structure observed at 350~\textmu m (see \figref{3color}).

\begin{table*}
 \caption{Properties of the Nine Superstructures}
  \label{tb:st9}
  \begin{tabular*}{18cm}
  {@{\extracolsep{\fill}}lccccccccc}
    \hline
    \hline
    \shortstack{ID} &
    \shortstack{Members\tablenotemark{\blue{a}}} &
    \shortstack{Radius\tablenotemark{\blue{b}}} &
    \shortstack{$\mu_{\alpha, \mathrm{in}}$\tablenotemark{\blue{c}}} &
    \shortstack{$S_{\mu_{\alpha, \mathrm{in}}}$\tablenotemark{\blue{d}}} &
    \shortstack{$\mu_{\delta, \mathrm{in}}$\tablenotemark{\blue{e}}} &
    \shortstack{$S_{\mu_{\delta,\mathrm{in}}}$\tablenotemark{\blue{f}}} &
    \shortstack{RV\tablenotemark{\blue{g}}} &
    \shortstack{$S_\mathrm{RV}$\tablenotemark{\blue{h}}} & \\
     & 
     &
     \colhead{ (pc) } &
     \colhead{ ($\mathrm{km~s^{-1}}$) } &
     \colhead{ ($\mathrm{km~s^{-1}}$) } &
     \colhead{($\mathrm{km~s^{-1}}$)} &
     \colhead{($\mathrm{km~s^{-1}}$)} &
     \colhead{($\mathrm{km~s^{-1}}$)} &
     \colhead{($\mathrm{km~s^{-1}}$)} & \\
    \hline
1& 52 (2)    &  61.8 & 52.70   & 41.19 & -13.00 & 30.95 & 176.4 & 7.1 \\
2& 158 (3)   & 106.4 & 55.27   & 42.02 & -26.59 & 24.46 & 169.4 & 7.9 \\
3& 187 (7)   & 101.4 & 25.89   & 42.20 & -12.06 & 40.66 & 177.0 & 7.4 \\
4& 234 (17)  &  93.7 & -10.17  & 28.83 & -12.53 & 32.16 & 174.4 & 16.4 \\
5& 560 (23)  & 154.4 & 42.74   & 49.04 & -2.18  & 38.39 & 163.8 & 31.7 \\
6& 1441 (79) & 172.5 & -15.66  & 49.86 &  2.65  & 33.69 & 160.1 & 18.6 \\
7& 896 (84)  & 202.5 & -8.08   & 39.78 &  0.38  & 33.32 & 144.9 & 10.8 \\
8& 145 (11)  &  73.5 & -15.91  & 72.09 &  4.72  & 48.00 & 127.6 & 21.3 \\
9& 257 (8)   & 104.5 & -34.81  & 43.82 & 10.84  & 31.63 & 140.6 & 14.7 \\

\hline
  \end{tabular*}
  \tablenotetext{\blue{a}}{The number of massive star candidates constituting the superstructures. The number of the clipped samples included in the superstructures, for which RV data are available, is denoted in parentheses.}
  \tablenotetext{\blue{b}}{The half-number radius calculated as the geometric mean of the semi-major and semi-minor axes obtained from elliptical fitting.}
  \tablenotetext{\blue{c}}{The median of $\mu_{\alpha, \mathrm{in}}$ for the clipped samples included in the superstructures, assuming a distance of 65~kpc. The calculation of the median does not account for coordinate distortion.}
  \tablenotetext{\blue{d}}{The standard deviation of $\mu_{\alpha,\mathrm{in}}$ for the clipped samples included in the superstructures.}
  \tablenotetext{\blue{e}}{The median of $\mu_{\delta,\mathrm{in}}$ for the clipped samples included in the superstructures, assuming a distance of 65~kpc. The calculation of the median does not account for coordinate distortion.}
  \tablenotetext{\blue{f}}{The standard deviation of $\mu_{\delta,\mathrm{in}}$ for the clipped samples included in the superstructures.}
  \tablenotetext{\blue{g}}{The mean RV of the clipped samples included in the superstructures, which is referenced to the solar system barycentric frame.}
  \tablenotetext{\blue{h}}{The standard deviation of RV for the clipped samples included in the superstructures.}
\end{table*}

\subsection{Three-dimensional Motions of the Superstructures}
\label{sec:RV}
In this section, we discuss the three-dimensional motions of the superstructures using the median internal PMs and mean RVs of the clipped samples that comprise the superstructures.
\figref{SMC_st9_PM} shows the median internal PMs and mean RVs\footnote{The reason for not employing the median as the representative value of RV is that, in superstructures where three or fewer stars have available RV, the median is not meaningful. Even if the medians were used as the representative value for RVs, the difference from the mean RVs is less than one-third of the standard deviations shown in \tabref{tb:st9}.} of the superstructures. The mean RVs shown in \figref{SMC_st9_PM} have been adjusted by subtracting the approximate systemic RV of the SMC, which is $160~\mathrm{km~s^{-1}}$ (e.g., \citealp{2004ApJ...604..176S, 2019MNRAS.483..392D}). \tabref{tb:st9} summarizes the properties of the nine superstructures, assuming a distance of 65~kpc. 
Particularly in the smaller superstructures within the wing, attention should be drawn to the limited number of clipped samples for which RVs are available.

As an overall trend, the superstructures in the eastern part of the SMC are moving eastward and receding, while those in the western part are moving westward and approaching, indicating that the eastern and western regions of the SMC are exhibiting opposite motions that are diverging from each other. The absolute values of median internal PMs and mean RVs with 160~$\mathrm{km~s^{-1}}$ subtracted are of similar magnitude. Although the standard deviations of the PMs for the superstructures are large, the overall direction of motion of the structure aligns with that shown in \figref{SMC_inPM}, suggests that the median internal PMs is significant.
A noteworthy point is that the superstructures~6, 5, and 4, corresponding to North~1, 2, and 3 discussed in \secref{sec:spacial}, are moving in different directions. The superstructure~6, which contains the largest number of massive star candidates among the superstructures, has a mean RV that is nearly stationary, while its median internal PM indicates movement to the west, in conjunction with superstructures belonging to the western bar. In contrast, the adjacent superstructure~5 has a median internal PM moving east alongside superstructures in the wing, while its mean RV remains nearly stationary.
Assuming that the distance along the line of sight is the same, the superstructures~6 and 5 overlap at $\approx$8~Myr ago (6.9 in the common logarithm), corresponding to the period of most intense star formation in the SMC \citep{2018MNRAS.478.5017R}.
The superstructure~4 exhibits an even larger mean RV but has a median internal PM directed westward.
\citet{2024AJ....168..255H} referred to the young stellar structures of the superstructure~4 and 5 as a shell-like structure, showing that they are moving in the RA direction and concluding that this motion is influenced by tidal interactions with the LMC. While the eastward motion of the superstructure~5 is consistent with the findings of \citet{2024AJ....168..255H}, we emphasize that the superstructure~4 moves in a direction clearly different from this.
As discussed in \secref{sec:pm}, different distances may result in different directions of the PM vectors. Therefore, the distinct motion directions of superstructures~4, 5, and 6 may partially originate from their significantly different line-of-sight depths. In any case, although the parameters used for identifying the superstructures in \secref{sec:identification_st} do not have physical meaning, it is clear that the superstructures possess not only spatial but also kinematic differences, indicating they are physically distinct structures.

The direction of the PMs of the superstructures may change when their respective distances are considered in the calculations.
Therefore, we apply the distance calculation method suggested by \citet{2016ApJS..224...21R, 2017MNRAS.472..808R} to the classical Cepheids in the SMC from the fourth phase of the Optical Gravitational Lensing Experiment (OGLE-IV survey; \citealp{2015AcA....65....1U, 2015AcA....65..297S}), and calculate the mean distances of the classical Cepheids overlapping each of the superstructures under the assumption of $D_\mathrm{SMC}=65~\mathrm{kpc}$.
Since few classical Cepheids are distributed in the wings, the calculations are performed only for the superstructures 4--9 located in the bar.
The mean distance of classical Cepheids within the 2$\sigma$ ellipses of the superstructures~4--9 is $\mathrm{\sim}61\text{--}67~\mathrm{kpc}$, with no structure significantly deviating from our assumed $D_\mathrm{SMC}=65~\mathrm{kpc}$.
Moreover, the classical Cepheids overlapping the superstructures~7--9 exhibited mean internal PMs directed to the northwest, when their distances are considered in the calculations.
This supports our hypothesis that the opposite motions seen in the east and west of the superstructures are not due to a projection of distances.
However, internal PMs pointing in different directions are not observed in classical Cepheids in the superstructures~4, 5, and 6.
This suggests that the spatial distributions and motions of our massive star candidates differ from those of the classical Cepheids, and attention should be paid to the fact that the massive star candidates are not necessarily centrally concentrated at $D_\mathrm{SMC}=65~\mathrm{kpc}$.
A precise discussion would require distance measurements for the massive star candidates, which we currently lack.
A more detailed comparison of the motions of the classical Cepheids and our massive star candidates will be addressed in a future paper.

\subsection{Young Stars without Galactic Rotation}
In this section, we discuss the origin of the opposite motions observed in the east and west of the SMC, as seen in \figref{SMC_st9_PM}.
Historically, the northeast-southwest symmetric velocity structure observed in H\,\textsc{i} has been interpreted as galactic rotation, and successful fittings of rotation models have been conducted \citep{2004ApJ...604..176S, 2019MNRAS.483..392D}. Thus, the opposite RVs observed in the superstructures might, at first glance, be interpreted as galactic rotation.

Meanwhile, the RV gradient of our massive star candidates is in the northwest-southeast direction, which is orthogonal to the velocity gradient of H\,\textsc{i}, and exhibits velocities that contradict the rotation models of H\,\textsc{i}.
The velocity gradient in the northwest-southeast direction is also observed in young and old stars in previous studies where RVs were independently measured \citep{2008MNRAS.386..826E, 2014MNRAS.442.1663D}, and cannot be explained by a sample deficiency or systematic errors in our data.
Additionally, in the previous studies, large RVs highlighting the northwest-southeast RV gradient have been measured in the southern region of the SMC (between the wing and the southwestern bar), where our massive star candidates are not distributed. Evolved intermediate-mass star candidates, which are slightly fainter than the massive star candidates (i.e., $\gtrsim$14 mag; see \figref{CMD1}), are also distributed in the southern region of the SMC. The \textit{Gaia} RVs of these stars also exhibit large values (see Figure 5 of \citealt{2024ApJ...962..120M}). These observations exclude the possibility that the northwest-southeast RV gradient is an apparent gradient caused by the wing, which is physically separated from the main body of the SMC.
At the same time, our massive star candidates are almost absent in the northeast, where the high-velocity end of the northeast-southwest H\,\textsc{i} velocity gradient is observed. 
A few massive star candidates that are slightly distributed in the northeast have small RVs (around 100~$\mathrm{km~s^{-1}}$), which do not match the H\,\textsc{i} redshifted component (around 200~$\mathrm{km~s^{-1}}$). The small RVs of the stars in the northeast are consistent with old red giants \citep{2014MNRAS.442.1663D}, further highlighting the contradiction with the galactic rotation suggested by H\,\textsc{i}.
Taking these facts into consideration, it is reasonable to conclude that the massive star candidates do not follow the galactic rotation as previously assumed.

Alternatively, the orthogonal velocity gradients of stars and H\,\textsc{i} may both arise from the same galactic rotation, with phenomena such as ram pressure, which affect only the gas, potentially causing a misalignment in the rotational axis.
An extreme phenomenon where ram pressure influences a galaxy is the gas stripping from member galaxies by the intracluster medium of galaxy clusters \citep{1972ApJ...176....1G}. However, even in these galaxy clusters, only a slight deviation from galactic rotation is observed in the gas of galaxies experiencing gas stripping (e.g., \citealp{2020ApJ...901...95C}). Therefore, it is unlikely that ram pressure has caused a significant misalignment of the gas rotation axis by as much as $\approx$90$^{\circ}$ in the SMC, even when considering the influence of the high-temperature gas associated with the local group.
For similar reasons, there is little justification for adopting a model that creates an extreme discrepancy between gas and stars, such as one where a face-on gas disk, devoid of stars, is located at $\sim$60~kpc, while stars extends over $\sim$20--30~kpc.

Furthermore, the east-west opposite PMs observed in \figref{SMC_inPM} and \figref{SMC_st9_PM} are inconsistent with the notion of galactic rotation. Instead, it appears that the massive star candidates in the eastern SMC are moving southeast toward the LMC, while those in the western SMC are moving northwest away from it.
This depiction can be interpreted as evidence of the SMC being disruptively elongated due to its interaction with the LMC.
Again, we emphasize that the trends observed in the internal PMs of our massive star candidates are similar to those found in previous studies that calculated the internal PMs of the SMC and discussed its tidal disruption \citep{2018ApJ...864...55Z, 2018ApJ...867L...8O, 2019ApJ...887..267M, 2021MNRAS.502.2859N}.
If we consider the effects of the interaction with the LMC, the differing RVs in the northwest and southeast can be understood not as rotation, but rather as elongation in the northwest-southeast direction, specifically as movement away to the southeast and movement toward the northwest.
Thus, we argue that the east-west opposing PMs indicate that at least galactic rotation is not dominant in the motion of the SMC. 

The idea is also supported by the simulation of \citet{2012ApJ...750...36D}. \citet{2012ApJ...750...36D} modeled the SMC using a rotating disk and a spheroid, simulating the gravitational interactions between the SMC, LMC, and the Milky Way. Their results showed that the rotating disk reproduced the northeast-southwest velocity gradient observed in H\,\textsc{i}, while the spheroid produced a line-of-sight velocity gradient in the northwest-southeast direction with velocities of $\sim$$100\text{--}200~\mathrm{km~s^{-1}}$, due to the tidal forces from the LMC. The line-of-sight velocities from the spheroid are similar to the RVs of the massive star candidates, suggesting that the observed RVs can be explained in the absence of rotating disk.

If the massive star candidates trace the motion of the ISM, then the velocity structures observed in H\,\textsc{i} would similarly not represent galactic rotation. 
Therefore, the dark matter mass and total mass of the SMC derived from previous studies that assumed galactic rotation, as well as the numerical simulations based on those findings, may need to be reconsidered.
Note that, as a more detailed discussion, it might also be meaningful to compare the velocity field with high-resolution observations of CO, which traces denser gas.

The distance to the LMC is 50~kpc, making it closer to us than the SMC, but massive star candidates located near the LMC on the celestial sphere exhibit large RVs\footnote{Due to the angular separation between the SMC and LMC on the celestial sphere, the line-of-sight defined by the two objects is slightly different in direction.}.
Although it may seem unusual at first, the increase in line-of-sight velocity towards the southeast along the bridge from the SMC is also observed in H\,\textsc{i} \citep{2005A&A...432...45B}.
It can be explained in two ways: Stars moving toward the southeast, in the direction of the LMC, are likely influenced by the gravitational potential of the LMC, approaching its systemic RV ($\sim$$260~\mathrm{km~s^{-1}}$; \citealp{2002AJ....124.2639V}).
This idea is also supported by the fact that the simulation of \citet{2012ApJ...750...36D} reproduced the northwest-southeast velocity gradient solely from gravitational interactions.
Alternatively, they may be decelerating due to ram pressure from the gas associated with the LMC and/or the Magellanic Corona \citep{2022Natur.609..915K}, causing them to move closer to the systemic RV of the LMC.
Regarding the latter explanation of ram pressure, if the Magellanic Corona exists, it is likely that the influence comes from the gas of the LMC rather than that of the Milky Way.
Since the magnitude of the ram pressure depends on the density of the gas that exerts the ram pressure and the velocity of the SMC relative to the gas, the idea of ram pressure stripping caused by the gas associated with the LMC is supported by two factors: the density of the gas in the LMC, including its associated Magellanic Corona, increases as it gets closer to the LMC; the relative velocity of the SMC with respect to the LMC is directed away from the LMC, almost in the opposite direction of the wing \citep{2013ApJ...764..161K,2017MNRAS.466.4711B}.
Depending on the collision parameters, it is also possible that not only tidal forces but also ram pressure from the LMC's gas has stripped the gas from the SMC. This ram pressure stripping is consistent with the different spatial distribution of the SMC's ISM and our massive star candidates compared to older stars. Additionally, it suggests that the gas tail may be inducing star formation, akin to the jellyfish galaxies observed in galaxy clusters (e.g., \citealp{2021gcf2.confE..33N}).

\citet{2019ApJ...887..267M} calculated the \textit{Gaia} DR2 internal PMs of 143 O-type and B-type stars, primarily selected from E04 stars, whose RVs match with H\,\textsc{i} peak, and showed that their motion is inconsistent with the RVs and internal PMs expected from the rotation model of \citet{2019MNRAS.483..392D}.
Our 7,426 massive star candidates further strengthen the assertion of \citet{2019ApJ...887..267M} by extending it across the entire SMC region.
At the same time, we show that similar conclusions can be drawn using different selection methods for gas tracers; specifically, our massive star candidates do not include many late B-type stars from the E04 stars (see \secref{sec:B10CMD}), as we apply stricter mass and age constraints ($M_\mathrm{ini}\geq8M_\odot$; $Age\lesssim 50.0~\mathrm{Myr}$).
In contrast, \citet{2024ApJ...962..120M} notes evidence of galactic rotation after accounting for the east-west gradient in PMs caused by tidal forces, using red supergiants with $Age \lesssim \mathrm{100~Myr}$. However, our massive star candidates exhibit PM gradients not only in the east-west direction but also in the north-south direction (see \figref{SMC_inPM}), indicating differing properties. Even when subtracting the east-west PM gradients and/or the north-south PM gradients from our massive star candidates, no clear evidence of galactic rotation is observed, which is also evident from the large velocity dispersions of the superstructures shown in \tabref{tb:st9}.

\citet{2021MNRAS.502.2859N} similarly argued that the motion of young stars ($\sim$140~Myr; \citealp{2019MNRAS.490.1076E}) and old stars ($\sim$4~Gyr) can be explained by tidal stretching rather than galaxy rotation.
However, based on Cepheid distances \citep{2017MNRAS.472..808R}, they claimed that stars closer to us (located on the eastern side of the SMC such as the wing) tend to move towards the northeast, which contrasts with our observed southeast-northwest elongation. 
Since we focus on younger massive stars ($Age\lesssim\mathrm{50.0~Myr}$), our findings are complementary to those of \citet{2021MNRAS.502.2859N}.
While our massive star candidates do exhibit PMs directed towards the outskirts, there is no significant flow in the northeast direction (see \figref{SMC_inPM}).
Our massive star candidates are localized near the northern H\,\textsc{i} center rather than in the bar where the optical center and Cepheid center are located (see \figref{SMC_PM}), suggesting their distances and motions should differ from those of the Cepheids.
Nevertheless, the idea that the redshifted component in the eastern SMC indicates stripping rather than rotation is a view we share.
Note that, since we have not determined the distances to the massive star candidates, we cannot rule out the possibility that nearby and distant massive star candidates may exhibit a southwest-northeast elongation. On the other hand, as mentioned in \secref{sec:RV}, we have confirmed that classical Cepheids in the southwestern part of the SMC tend to move toward the northwest, and it is also worth noting that a few classical Cepheids near the wing are moving toward the southeast.

Additionally, \citet{2021MNRAS.502.2859N} showed that the stars in the northwestern part of the SMC are moving toward the northwest, and discussed that this motion may correspond to the ``Counter-Bridge" — a tidal arm, as inferred from the simulations of \citet{2012ApJ...750...36D}, which extends from the western part of the SMC, stretches behind the SMC, and arcs toward the LMC.
As we have mentioned in \secref{sec:identification_st} and shown in \figref{st7_nw}, several lines of massive star candidates extend toward the northwest from the superstructure~7. This can corresponds to the northwestward flow discovered by \citet{2021MNRAS.502.2859N}.
However, as shown in \figref{SMC_RV}, RVs of the few stars in the northwestern part of the SMC are slightly larger than the smaller RVs in the bar structure, reaching up to $\lesssim$150~$\mathrm{km~s^{-1}}$. There is no evidence supporting large RVs directed behind the SMC, and this holds true even when including all massive star candidates, except for the clipped samples.
Therefore, while there is no strong evidence that these stars moving toward the northwest correspond to the Counter-Bridge proposed by \citet{2012ApJ...750...36D}\footnote{The simulation suggests that the Counter-Bridge was formed simultaneously with the Magellanic Bridge, and it is reasonable that the Counter-Bridge is not observed in classical Cepheids, which do not distribute in the wing and bridge regions. At the same time, the idea of explaining the $\sim$20--30~kpc line-of-sight depth observed in classical Cepheids solely by the Counter-Bridge may not be reasonable.}, it is likely that they represent a tidal feature moving in the opposite direction to the group of stars moving toward the southeast.

The reason previous studies successfully fitted rotation models for H\,\textsc{i} may lie in a misinterpretation of the deep structures along the line-of-sight of the SMC.
There are also simulation results suggesting that the redshifted component in the northeast may correspond to the projection of the Counter-Bridge behind the SMC, which is located at $\sim$80~kpc \citep{2007MNRAS.381L..11M}.
While accurately determining the distance to the gas is challenging, distances to stars can be measured more reliably.
For example, \citet{2024ApJ...962..120M} has claimed the presence of a bimodal distribution at distances of 61~kpc and 66~kpc for red supergiants with $Age\lesssim\mathrm{100~Myr}$, based on their estimates of extinction.
We are exploring the use of isochrone fitting to determine the distances of our massive star candidates, and we plan to discuss this in a future paper (Nakano et al. in prep).

In contrast to the SMC, the internal PMs of the LMC, after subtracting the systemic PM, are dominated by galactic rotation \citep{2021A&A...649A...7G}. Furthermore, we have applied a similar methodology for selecting massive star candidates to the LMC (Tachihara et al. in prep), and the PMs of the LMC's massive star candidates, after subtracting the systemic PM, are also found to be dominated by galactic rotation. This presence of galactic rotation in the LMC further emphasizes the absence of galactic rotation in the SMC.

\section{Summary and Conclusions}
\label{sec:conclusion}
In this study, we selected non-embedded 7,426 massive star candidates ($M_\mathrm{ini}\geq8M_\odot$, $Age\lesssim\mathrm{50.0~Myr}$) based on the uniform observations with \textit{Gaia}. Our catalog represents the first homogeneous catalog of massive stars in the SMC and contains the largest sample size. The massive star candidates are spatially aligned with the distribution of the ISM and exhibit motion consistent with that of younger stars in prior studies, indicating the validity of our selection process.
The PMs of the massive star candidates predominantly point towards the outer regions of the SMC, with some objects moving at $\sim$100~$\mathrm{km~s^{-1}}$, suggesting a potential escape from the SMC. This phenomenon may correspond to the mass loss processes that the SMC undergoes due to interactions and feedback mechanisms.
We separated the galaxy into nine superstructures on a scale of $\sim$100~pc based on surface density, confirming that each possesses distinct kinematic properties. The superstructures exhibit opposite PMs and RVs in the east-west direction, moving away from each other; the eastern region of the SMC moves southeast with a higher line-of-sight velocity, while the western region moves northwest with a lower line-of-sight velocity. This opposite motion is consistent with the influence of interactions with the LMC, which located to the southwest, and can be explained by tidal forces and/or ram pressure, thereby rejecting the presence of galaxy rotation in the SMC.

If galactic rotation has been misidentified even in the SMC, this suggests that claims regarding the existence of galactic rotation in distant galaxies should also give significant attention to the presence of surrounding companions.
Also, if galaxy rotation is absent in the SMC, it could significantly alter the previously calculated histories of interactions among the Milky Way, LMC, and SMC.
To facilitate a more precise discussion, it is essential to determine the distances to the massive star candidates.
By elucidating the three-dimensional spatial and velocity distributions of young stars that trace the ISM, we can engage in a detailed examination of the dynamics of the SMC.
We plan to determine the distances to the massive star candidates through isochrone fitting of star clusters, and we have already identified several clusters containing many massive star candidates plotted on the isochrone (Nakano et al. in prep). Furthermore, once we obtain the three-dimensional spatial and velocity distributions of young stars, we can conduct simulations to replicate these distributions. By elucidating the massive star formation processes in the SMC through simulations, we aim to establish the SMC as a template for massive star formation in low-metallicity environments and in the context of galaxy interactions.

\begin{acknowledgments}
This work was financially supported by JST SPRING, Grant Number
JPMJSP2125. The author SN would like to take this opportunity to thank the
``ATHERS Make New Standards Program for the Next Generation Researchers."
KT is supported in part by JSPS KAKENHI Grant Number 20H01945.
We would like to thank Dr. Kazuki Tokuda for the insightful discussions regarding the ALMA data.
SN would like to acknowledge the language assistance provided by ChatGPT (\url{https://chatgpt.com/}) and DeepL (\url{https://www.deepl.com/}) during the preparation of this manuscript. ChatGPT was particularly helpful in refining English expressions and enhancing the clarity of the text, while DeepL contributed to accurate and fluent translations. 
This work has made use of data from the European Space Agency (ESA) mission
{\it Gaia} (\url{https://www.cosmos.esa.int/gaia}), processed by the {\it Gaia}
Data Processing and Analysis Consortium (DPAC,
\url{https://www.cosmos.esa.int/web/gaia/dpac/consortium}). Funding for the DPAC
has been provided by national institutions, in particular the institutions
participating in the {\it Gaia} Multilateral Agreement.
This research has made use of the VizieR catalogue access tool, CDS,
Strasbourg, France \citep{10.26093/cds/vizier}. The original description 
of the VizieR service was published in \citet{vizier2000}.
Cerro Tololo Inter-American Observatory (CTIO) is operated by the Association of Universities for Research in Astronomy Inc. (AURA), under a cooperative agreement with the National Science Foundation (NSF) as part of the National Optical Astronomy Observatories (NOAO). The MCELS is funded through the support of the Dean B. McLaughlin fund at the University of Michigan and through NSF grant 9540747.
\textit{Herschel} is an ESA space observatory with science instruments provided by European-led Principal Investigator consortia and with important participation from NASA.
This scientific work uses data obtained from Inyarrimanha Ilgari Bundara, the CSIRO Murchison Radio-astronomy Observatory. We acknowledge the Wajarri Yamaji People as the Traditional Owners and native title holders of the Observatory site. CSIRO’s ASKAP radio telescope and Murriyang, CSIRO's Parkes radio telescope, are part of the Australia Telescope National Facility (\url{https://ror.org/05qajvd42}). Operation of ASKAP is funded by the Australian Government with support from the National Collaborative Research Infrastructure Strategy. Parkes is funded by the Australian Government for operation as a National Facility managed by CSIRO. ASKAP uses the resources of the Pawsey Supercomputing Research Centre. Establishment of ASKAP, Inyarrimanha Ilgari Bundara, the CSIRO Murchison Radio-astronomy Observatory and the Pawsey Supercomputing Research Centre are initiatives of the Australian Government, with support from the Government of Western Australia and the Science and Industry Endowment Fund.
This paper makes use of the following ALMA data: ADS/JAO.ALMA\#2017.A.00054.S, ADS/JAO.ALMA\#2018.1.01115.S, and ADS/JAO.ALMA\#2018.1.01319.S. ALMA is a partnership of ESO (representing its member states), NSF (USA) and NINS (Japan), together with NRC (Canada), MOST and ASIAA (Taiwan), and KASI (Republic of Korea), in cooperation with the Republic of Chile.
The Joint ALMA Observatory is operated by ESO, AUI/NRAO, and NAOJ.
.......
\end{acknowledgments}

%

\vspace{5mm}
\facilities{ALMA, ASKAP, \textit{Gaia}, \textit{Herschel}, Parkes}


\software{aplpy \citep{2012ascl.soft08017R}, astropy \citep{2013A&A...558A..33A,2018AJ....156..123A}, CASA \citep{2007ASPC..376..127M}, ChatGPT \citep{2024arXiv241021276O}, DBSCAN \citep{1996kddm.conf..226E}, dustapprox \citep{Fouesneau_dustapprox_2022}, matplotlib \citep{2007CSE.....9...90H}
}

\appendix

\section{Evaluation of the number of foreground objects belong to the Milky Way}
\label{sec:foreground}

\begin{figure*}
 \centering
 \includegraphics[width=18cm,clip]{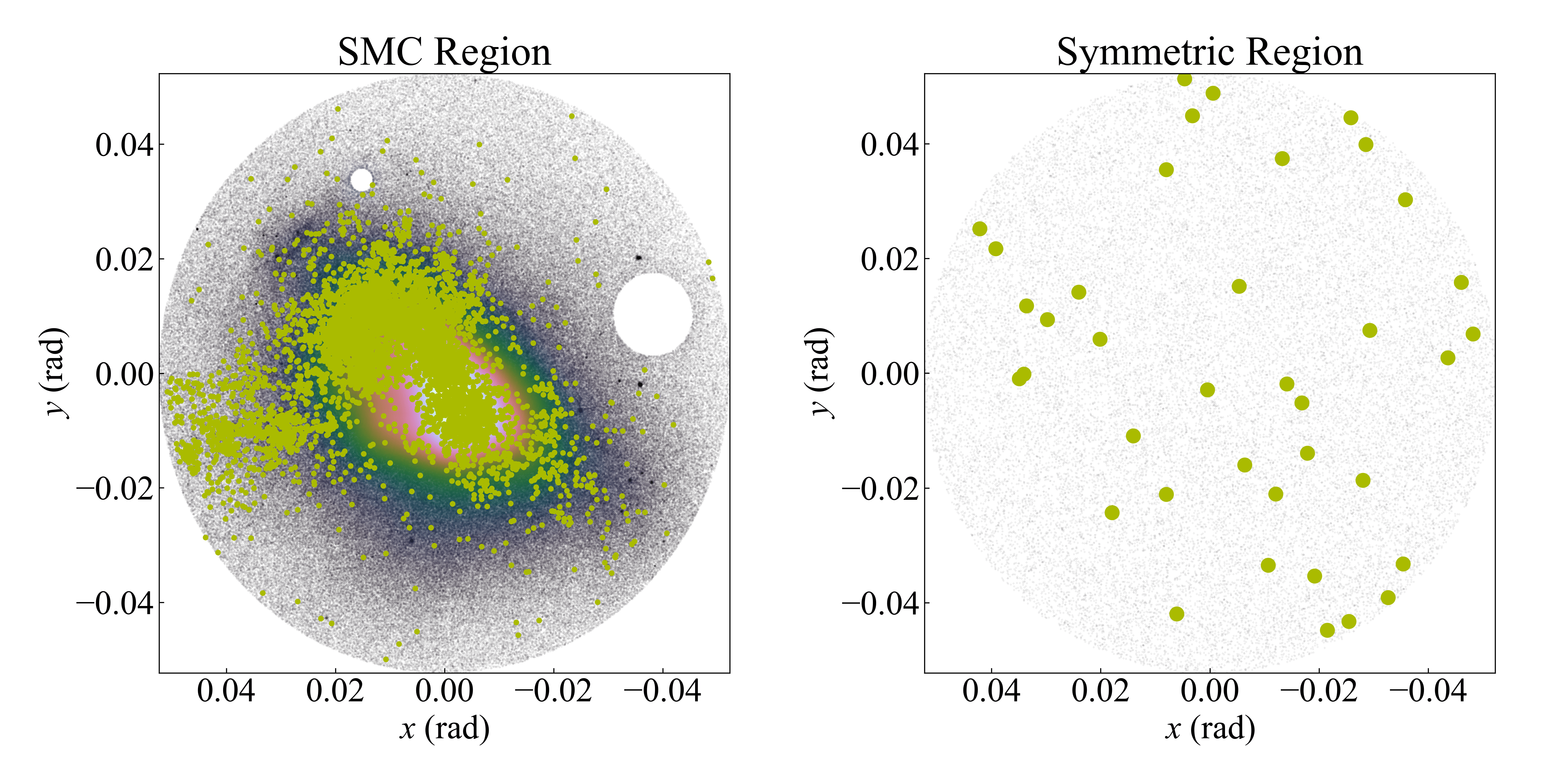}
 \caption{
 (Left) Stellar density distribution of the newly defined SMC region, contained within a circle of radius $3^{\circ}$ centered at $l = 302.8084^{\circ}, b = -44.3277^{\circ}$. Foreground objects are excluded using the restriction $\varpi \leq 0.1~\mathrm{mas}$ and cuts around the two globular clusters NGC~362 and NGC~104. Positions are shown in the ($x, y$) coordinates based on the orthographic projection defined in \secref{sec:pm}. The overplotted dark yellow circles represent the stars plotted above the magenta line in \figref{sym_CMD}. The newly defined circular SMC region contains 6,828 dark yellow circles. 
 (Right) Stellar density distribution of the region symmetric to the SMC region relative to the galactic plane, contained within a circle of radius $3^{\circ}$ centered at $l = 302.8084^{\circ}, b = +44.3277^{\circ}$. Foreground objects are excluded using the restriction $\varpi \leq0.1~\mathrm{mas}$. Positions are shown in the ($x, y$) coordinates based on the orthographic projection defined in \secref{sec:pm}. The overplotted dark yellow circles represent the stars plotted above the magenta line in \figref{sym_CMD}. The symmetric region contains 38 dark yellow circles. The dark yellow circles in the symmetric region are enlarged for visibility. 
 }
\label{sym_plx}
\end{figure*}

In this section, we evaluate the number of foreground objects contained within the SMC region defined in \secref{sec:data1}. The evaluation strategy involves obtaining and comparing \textit{Gaia} DR3 data from a region symmetric to the SMC region relative to the galactic plane, where only objects from the Milky Way are expected to exist. After applying the same criterion $\varpi\leq0.1~\mathrm{mas}$ to the data from the symmetric region as that applied to the SMC region, we count the number of stars plotted in the region where massive stars exist on the CMD defined in \secref{sec:CMD}.
Since the symmetric region should contain only objects from the Milky Way, the number of stars plotted in the CMD massive star area corresponds entirely to foreground objects. This count is estimated to be approximately equal to the number of foreground objects in the SMC region, assuming the area of the symmetric region is defined to be equal to that of the SMC region.

For a fair comparison, it is necessary to ensure that the areas of the SMC region and the symmetric region are equal.
In this study, the SMC region was defined using four specified points of RA and DEC ($\mathrm{RA}=00\mathrm{h}\textsc{--}03\mathrm{h}$ and $\mathrm{DEC}=-76^{\circ} \mathrm{\,\,to\,\,} -69^{\circ}$). To facilitate the acquisition of data with equal area, this section redefines the SMC region as a circle with a radius of $3^{\circ}$ centered on point $\mathrm{RA}=00\mathrm{h}\,\,52\mathrm{m}\,\,37.99\mathrm{s}$, $\mathrm{DEC}=-72^{\circ}\,\,48'\,\,01.08''$ (which correspond to $l = 302.8084^{\circ}, b = -44.3277^{\circ}$ in galactic coordinates). As in \secref{sec:data1}, two globular clusters is removed from this new SMC region.
The data for the region symmetric to the SMC region relative to the galactic plane are obtained by reversing the sign of the galactic latitude of the center coordinates of the circle. Specifically, a circle with a radius of $3^{\circ}$ centered at $l = 302.8084^{\circ}, b = +44.3277^{\circ}$ is defined as the symmetric region. Although the area of the symmetric region is slightly larger because it does not exclude globular clusters, this results in a higher estimate of foreground objects that may contaminate the SMC region, which does not pose an issue for demonstrating a lower contamination of foreground objects.
\figref{sym_plx} shows the stellar density distribution created by applying the condition $\varpi \leq 0.1~\mathrm{mas}$ to the \textit{Gaia} DR3 data for the defined SMC region and the symmetric region. The new SMC region differs only in extent from \figref{SMC_Gaia}, with stars exhibiting a similar spatial distribution that clearly reveals the shape of the SMC. This new SMC region contains 1,834,892 stars. In contrast, the spatial distribution of stars in the symmetric region appears featureless, indicating that the plotted stars are likely from the Milky Way.
The symmetric region contains 32,105 stars. At this point, the number of stars in the symmetric region is less than 2\% of the stars in the circular SMC region, suggesting that the proportion of foreground objects in the SMC region is likely to be small.

\begin{figure}
 \centering
 \includegraphics[width=8.5cm,clip]{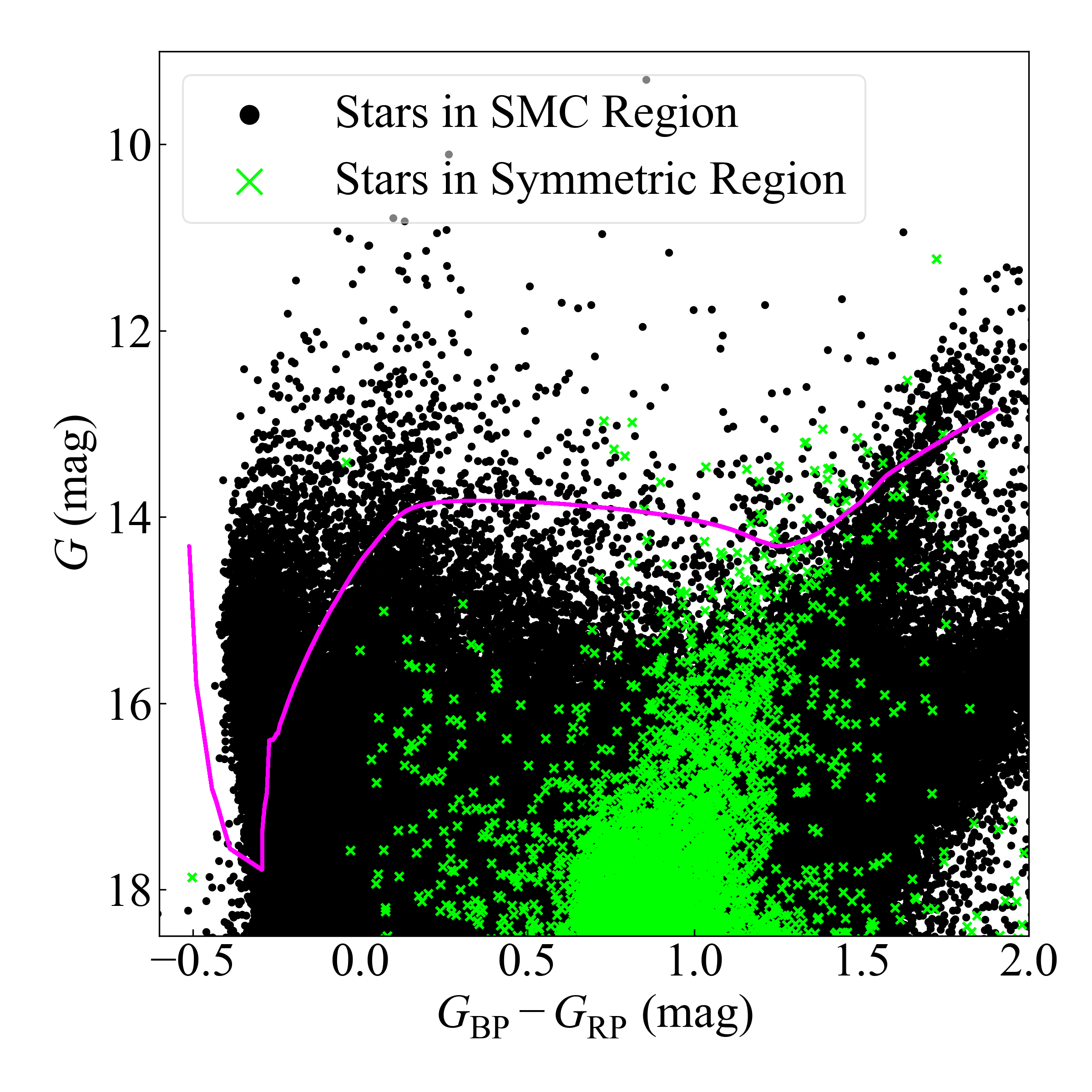}
 \caption{The ($G_\mathrm{BP}-G_\mathrm{RP}$, $G$) CMD after excluding foreground objects from the newly defined circular SMC region and the symmetric region with respect to the galactic plane. The region above the magenta line is the area occupied by massive stars in the CMD, defined in \secref{sec:data1}. Stars contained within the SMC region are represented by black circles, while stars from the symmetric region are represented by green crosses.
 }
\label{sym_CMD}
\end{figure}

\figref{sym_CMD} shows ($G_\mathrm{BP}-G_\mathrm{RP}$, $G$) CMD of the stars plotted in \figref{sym_plx}. The region above the magenta line represents the area where massive stars are expected, and the newly defined circular SMC region contains 6,828 massive star candidates. In contrast, the stars in the symmetric region also include some that fall within the massive star region, totaling 38.
The dark yellow circles of \figref{sym_plx} shows the spatial distribution of stars plotted in the massive star region of \figref{sym_CMD}. The spatial distribution of massive star candidates in the SMC region exhibits the bar structure and wing structure, similar to what is shown in \figref{SMC_Gaia_withOB}. In contrast, the spatial distribution of stars in the massive star region of the symmetric region appears to be randomly scattered, suggesting that these stars are more likely to be misidentified low- to intermediate-mass stars from the Milky Way rather than selected massive stars.
If foreground objects from the Milky Way are present in the SMC region at the same ratio as in the symmetric region, it is expected that an equivalent number of foreground objects are included in the newly defined circular SMC region. This means that out of the 6,828 massive star candidates, 38 are likely foreground objects belonging to the Milky Way, resulting in an estimated false detection rate of approximately 0.56\%.
We note that in \figref{sym_CMD}, if we focus on blue stars with $G_\mathrm{BP}-G_\mathrm{RP}\leq0.5~\mathrm{mag}$, we can narrow down the candidates to a single star in the symmetric region while potentially missing the SMC's yellow supergiants and red supergiants. This adjustment reduces the estimated false detection rate to approximately 0.02\%. We consider the $\sim$0.56\% false detection rate statistically negligible and opted not to apply a stronger restriction on $G_\mathrm{BP}-G_\mathrm{RP}$. However, if a cleaner sample is desired, it is recommended to impose a stronger restriction on color.

\section{Distribution on ($G_\mathrm{BP}-G_\mathrm{RP}$, $G$) CMD of various star populations in the B10 catalog}
\label{sec:B10CMD}

\begin{figure*}
 \centering
 \includegraphics[width=18cm,clip]{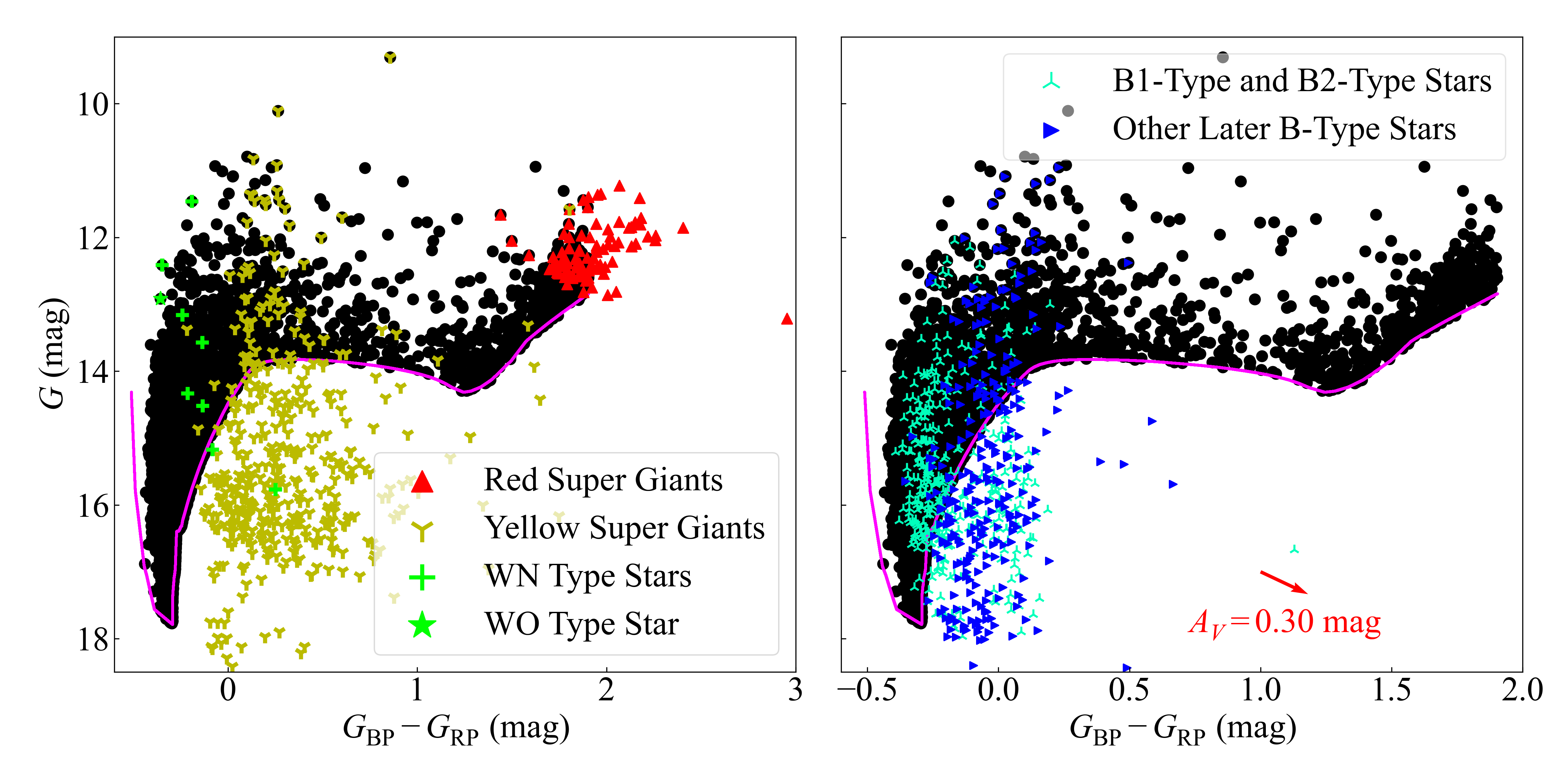}
 \caption{
  (Left) The ($G_\mathrm{BP}-G_\mathrm{RP}$, $G$) CMD of cross-matched red supergiants, yellow supergiants, and Wolf-Rayet stars from the B10 catalog. The 87 red supergiants are represented by red triangles, the 441 yellow supergiants by yellow Y-shaped symbols, the 8 WN-type stars by green crosses, and the single WO-type star by a green star symbol. The black circles represent the 7,426 massive star candidates identified in this study. The area above the magenta line represents the selected region for massive stars.
  (Right) The ($G_\mathrm{BP}-G_\mathrm{RP}$, $G$) CMD of cross-matched B1-type stars, B2-type stars, and other later B-type stars (B3--B9) from the B10 catalog. The 186 B1-type and B2-type stars are represented by cyan Y-shaped symbols, while the other later B-type stars are represented by blue triangles. The black circles represent the 7,426 massive star candidates identified in this study. The area above the magenta line represents the selected region for massive stars. The red arrow represents the reddening vector corresponding to $A_{V}=0.30~\mathrm{mag}$.
}
\label{B10_CMD_Other}
\end{figure*}

In \secref{sec:bonanos}, the discussion focused on B0-type and O-type stars, while this section presents the distribution of other stellar populations on the ($G_\mathrm{BP}-G_\mathrm{RP}$, $G$) CMD.
The left panel of \figref{B10_CMD_Other} shows the distribution on the CMD of red supergiants, yellow supergiants, and Wolf-Rayet stars from the B10 catalog, cross-matched in \textit{Gaia} DR3 data. For this analysis, red supergiants are defined as stars containing either ``K" or ``M" in the classification string provided by the B10 catalog, while yellow supergiants are defined as those containing ``A," ``F," or ``G."

Red supergiants from the B10 catalog include stars with $G_\mathrm{BP}-G_\mathrm{RP}\geq1.9$, which exceed our selection range; however, more than half (47 out of 87) are included among our massive star candidates. If we allow for confusion with AGB stars, extending the massive star detection region to $G_\mathrm{BP}-G_\mathrm{RP}= 2.5~\mathrm{mag}$ would enable us to capture nearly all the red supergiants from the B10 catalog. This suggests that our selection method is robust and reliable.

Yellow supergiants from the B10 catalog comprise only 71 out of 441 in our massive star candidates, accounting for a large proportion of the stars in the catalog that are not included in our selection. Those located near the boundary of the massive star selection area on the CMD, as shown in \figref{B10_CMD_Other}, may be excluded due to reddening caused by circumstellar dust. However, if spectroscopically identified yellow supergiants, which should inherently be ``yellow," have initial masses of $M_\mathrm{ini}\geq8M_{\odot}$, they must be sufficiently bright with $G\geq14~\mathrm{mag}$ before experiencing any reddening.
Despite this, most yellow supergiants that fall outside the massive star selection area in \figref{B10_CMD_Other} would shift to a relatively dim ``blue" region with $G_\mathrm{BP}-G_\mathrm{RP}\leq0$ when corrected for dereddening to enter the selection area.
This fact contradicts spectral classifications from the B10 catalog, leading us to conclude that the yellow supergiants not included in our massive star candidates likely comprise a large number of intermediate-mass stars. This underscores the necessity of focusing on B0-type and O-type stars in our discussions, as highlighted in \secref{sec:bonanos}.

In the SMC, 12 Wolf-Rayet stars have been identified \citep{2003PASP..115.1265M}, and subsequent observations have not revealed any new objects, leading to the conclusion that the population of Wolf-Rayet stars in the SMC is complete \citep{2018ApJ...863..181N}.
The B10 catalog includes all 12 Wolf-Rayet stars, but successful cross-matching with \textit{Gaia} DR3 is achieved for only 9 of them. Among these, 7 are included in our selected massive star candidates.
Wolf-Rayet stars likely experience reddening due to circumstellar dust; however, their detection rate remains high. Additionally, the only WO-type star in the SMC is included in our massive star candidates.

The right panel of \figref{B10_CMD_Other} shows the distribution on the CMD of stars classified as B1-type, B2-type, and other later B-type stars among those successfully cross-matched with the \textit{Gaia} DR3 and the B10 catalog.
To reduce the impact of uncertainties in spectral classification, we removed stars with broad spectral types from the B10 catalog.
The category of B1-type and B2-type stars do not include stars that could be classified as B3-type or later, as well as those that could be classified as B0-type.
Similarly, the category of other later B-type stars do not include stars that could be classified as B2-type stars or earlier, as well as those that could be classified as A-type.

B1-type stars and B2-type stars from the B10 catalog that are included in our massive star candidates are 63 out of 103 and 19 out of 83, respectively.
Although these percentages are lower compared to B0-type and O-type stars, B1-type and B2-type stars are localized around the dust-rich NGC~346 region, making them dependent on specific observations. Therefore, we exclude B1-type and B2-type stars from the discussion in \secref{sec:B10OB0}.
Other B3--B9 type stars are predominantly composed of evolved giants, but similar to yellow supergiants, they are likely to contain a large number of intermediate-mass stars for analogous reasons.


\bibliographystyle{splncs04}



\end{document}